\documentclass[aps,floats]{revtex4}
\usepackage{amsmath,amssymb}
\usepackage{graphicx,epsfig}

\begin{document}
\bibliographystyle {plain}

\def\oppropto{\mathop{\propto}} 
\def\opsimeq{\mathop{\simeq}}
\def\opoverderline{\mathop{\overline}}
\def\operarrow{\mathop{\longrightarrow}}
\def\opsim{\mathop{\sim}}

\def\fig#1#2{\includegraphics[height=#1]{#2}}
\def\figx#1#2{\includegraphics[width=#1]{#2}}


\title{ Block Renormalization for quantum Ising models in dimension $d=2$ : \\
applications to the pure and random ferromagnet, and to the spin-glass } 


 \author{ C\'ecile Monthus}
  \affiliation{Institut de Physique Th\'{e}orique, 
CNRS and CEA Saclay, 
 91191 Gif-sur-Yvette cedex, France}

\begin{abstract}

For the quantum Ising chain, the self-dual block renormalization procedure of Fernandez-Pacheco [Phys. Rev. D 19, 3173 (1979)] is known to reproduce exactly the location of the zero-temperature critical point and the correlation length exponent $\nu=1$. Recently, Miyazaki and Nishimori [Phys. Rev. E 87, 032154 (2013)] have proposed to study the disordered quantum Ising model in dimensions $d>1$ by applying the Fernandez-Pacheco procedure successively in each direction. To avoid the inequivalence of directions of their approach, we propose here an alternative procedure where the $d$ directions are treated on the same footing. For the pure model, this leads to the correlation length exponents $\nu \simeq 0.625$ in $d=2$ (to be compared with the 3D classical Ising model exponent $\nu \simeq 0.63$) and $\nu \simeq 0.5018$ (to be compared with the 4D classical Ising model mean-field exponent $\nu =1/2$). For the disordered model in dimension $d=2$, either ferromagnetic or spin-glass, the numerical application of the renormalization rules to samples of linear size $L=4096$ yields that the transition is governed by an Infinite Disorder Fixed Point, with the activated exponent $\psi \simeq 0.65$, the typical correlation exponent $\nu_{typ} \simeq 0.44$ and the finite-size correlation exponent $\nu_{FS} \simeq 1.25$. We discuss the similarities and differences with the Strong Disorder Renormalization results.

\end{abstract}

\maketitle

\section{ Introduction }

\label{sec_intro}

The quantum Ising model defined in terms of Pauli matrices $(\sigma_i^x,\sigma_i^z)$
\begin{eqnarray}
H= - \sum_{i}  h_i \sigma^x_i - \sum_{<i,j>} J_{i,j} \sigma_i^z \sigma^z_j 
\label{hqising}
\end{eqnarray}
is the basic model to study quantum phase transitions at zero-temperature \cite{sachdev}.
On a hypercubic lattice in dimension $d$, the pure model with the same transverse field $h$ on all sites, and the same ferromagnetic coupling $J$ between nearest-neighbors is well understood
via the equivalence with the classical Ising model in dimension $d^{class}=d+1$,
i.e. the time plays the role of an extra space-dimension \cite{sachdev},
and the dynamical exponent is $z_{pure} = 1$.
In particular, the quantum model in $d=1$ corresponds to the exactly solved 
2D classical Ising model,
the quantum model in $d=2$ corresponds to the 3D classical Ising model,
and the quantum model in dimension $d \geq 3$ is characterized by the standard mean-field exponents.

 From the point of view of Block-Renormalization for quantum models, there exists a special self-dual procedure introduced by Fernandez-Pacheco  \cite{pacheco,igloiSD},
which is able to reproduce the exact critical point $(J/h)_c=1$ and the exact correlation length exponent $\nu(d=1)=\nu(d^{class}=2)=1$. Various generalizations of this procedure to higher dimensions $d>1$ have been studied \cite{epele,nishiPur,kubica}, as well as extensions to other quantum models like the Potts and the Ashkin-Teller models \cite{horn,hu,solyom,solyom2}.

In the disordered case, where the transverse fields $h_i$ and the couplings $J_{i,j}$ are random variables, many exact results have been obtained in $d=1$ by Daniel Fisher \cite{fisher}  via the asymptotically exact strong disorder renormalization procedure
 (for a review, see \cite{review_strong}). In particular, the transition is governed by an Infinite Disorder Fixed Point and presents unconventional scaling laws with respect to the pure case.
In dimension $d>1$, the strong disorder renormalization procedure  has been studied numerically with the conclusion that the transition is also governed by an Infinite-Disorder fixed point in dimensions $d=2,3,4$  \cite{motrunich,fisherreview,lin,karevski,lin07,yu,kovacsstrip,kovacs2d,kovacs3d,kovacsentropy,kovacsreview}. These numerical renormalization results are in agreement with the results of independent quantum Monte-Carlo in $d=2$ \cite{pich,rieger}.
The Strong Disorder Renormalization is thus a very powerful method, but
it leads to a complicated renormalized topology for the surviving clusters as soon as $d>1$. In particular, a large number of very weak bonds are a priori generated
during the RG, that will eventually not be important for the forthcoming RG steps.
This is why recent numerical implementations of Strong Disorder RG rules are based
on algorithms avoiding this proliferation of weak generated bonds \cite{kovacs2d,kovacs3d,kovacsentropy,kovacsreview}.

A natural question is whether Infinite Disorder Fixed Points can be also reproduced
by more standard block-renormalization.
Recently, Miyazaki and Nishimori \cite{nishiRandom} have proposed to generalize 
the block-renormalization of Fernandez-Pacheco \cite{pacheco} concerning the pure model in $d=1$, and their extension for the pure model in $d=2,3$ \cite{nishiPur}
 to the random case : their results are in agreement with the Infinite Disorder Fixed Point scalings. However in dimension $d>1$, their procedure has the drawback that the various directions are treated inequivalently, so that they need to re-symmetrize afterward \cite{nishiPur}. In addition, in the random case, they have used some pool method that does not keep all the generated correlations between renormalized parameters, and the only critical exponent that they measure is the finite-size correlation length exponent $\nu$, whereas it is interesting to measure also the activated exponent $\psi$ and the typical correlation length exponent $\nu_{typ}.$

In the present paper, we thus propose another generalization
 in dimensions $d=2$ and $d=3$ of 
the self-dual one-dimensional block-renormalization of Fernandez-Pacheco, that we study
analytically in the pure case and numerically in the disordered case.
The paper is organized as follows.
In section \ref{sec_rgrule}, we describe the elementary renormalization rule
that will be applied throughout the paper. The application to the one-dimensional
quantum Ising model of Eq. \ref{hqising}, both pure and random, is recalled in section
 \ref{sec_d1}. In section \ref{sec_d2}, we derive the Block Renormalization rules
for the two-dimensional case. The applications of these two-dimensional rules
 are discussed respectively
in section \ref{sec_d2pure} for the pure ferromagnetic model, in section \ref{sec_d2randomferro} for the random ferromagnetic model, and in section \ref{sec_d2spinglass} for the spin-glass model.
Our conclusions are summarized in section \ref{sec_conclusion}.
The extension to dimension $d=3$ is presented in Appendix \ref{sec_d3}.

\section{ Elementary renormalization rule }

\label{sec_rgrule}

In this section, we consider the following elementary Hamiltonian involving $(b+1)$ spins
$\sigma_{i} $ with $i=0,1,2,..,b$
\begin{eqnarray}
H_b=  \sum_{i=1}^b 
\left(- h_i \sigma^x_i - J_{0,i} \sigma_{0}^z \sigma^z_i  \right)
\label{hintra}
\end{eqnarray}
where $\sigma_{0} $ plays the special role of the 'master', and the $b$ spins 
$\sigma_{i} $ with $i=1,2,..,b$ are the 'slaves'.

\subsection{ Diagonalization of $H_b$ } 

For each eigenvalue $S_{0}=\pm 1$ of $\sigma_{0}^z$,
one has to diagonalize independently the $b$ Hamiltonians
concerning the single quantum spin $\sigma_{i} $ in the transverse field $h_i$
and in the effective magnetic field $( J_{0,i} S_{0})$
\begin{eqnarray}
h^{(S_0)}_{i}= - h_{i} \sigma_{i}^x - J_{0,i} S_{0} \sigma_{i}^z 
\label{hintraeffi}
\end{eqnarray}
The two eigenvalues read
\begin{eqnarray}
\lambda^{\pm}_{i}(S_{0})= \pm \sqrt{ h_{i}^2 +J_{0,i}^2  }
\label{blocklambda}
\end{eqnarray}
with the following corresponding eigenvectors
\begin{eqnarray}
\vert \lambda^{-}_{i}(S_{0}) > && = \frac{ \vert S_{i}=+1 >
+c_i(S_{0}) \vert S_{i}=-1 > }{\sqrt{1+c_i^2(S_{0})}}  
\nonumber \\
\vert \lambda^{+}_{i}(S_{0}) > && =  \frac{ -c_i(S_{0}) \vert S_{i}=+1 >
+ \vert S_{i}=-1 > }{\sqrt{1+c_i^2(S_{0})}}  
\label{blocklambda1eigen}
\end{eqnarray}
where 
\begin{eqnarray}
c_i(S_{0}) && = \frac{ h_{i}}{ \sqrt{ h_{i}^2 +J_{0,i}^2 }
 +J_{0,i} S_{0} } =\frac{ \sqrt{ h_{i}^2 +J_{0,i}^2 }
 -J_{0,i} S_{0} } { h_{i}} = \frac{1}{c_i(-S_{0})}
\label{theta}
\end{eqnarray}

In summary, the Hamiltonian $H_b$ of Eq. \ref{hintra} has two degenerate ground states 
labeled by the two values $S_{0}=\pm 1$.  
The ground-state energy 
is simply obtained by adding the $b$ contributions $\lambda^{-}_{i}(S_{0}) $
\begin{eqnarray}
E_{GS}^{(S_0)} =\sum_{i=1}^b \lambda^{-}_{i}(S_{0}) = - \sum_{i=1}^b \sqrt{ h_{i}^2 +J_{0,i}^2  } 
\label{egsintra}
\end{eqnarray}
The two corresponding
ground-states are obtained by the tensor products
\begin{eqnarray}
\vert {GS}^{(S_0)} > = \vert S_{0} > \otimes (\otimes_{i=1}^b \vert \lambda^{-}_{i}(S_{0}) > )
\label{GSintra}
\end{eqnarray}

\subsection{ Projection onto the two lowest states of $H_b$ } 

The projector onto the two ground-states reads
\begin{eqnarray}
P_b \equiv \sum_{S_{0}=\pm 1} \vert {GS}^{(S_0)}> <  {GS}^{(S_0)} \vert
\label{projGSintra}
\end{eqnarray}
It is thus convenient to define the renormalized spin $\sigma_{R0} $  
from these two ground-states 
\begin{eqnarray}
\vert \sigma_{R0}^z=S_{0}> \equiv \vert {GS}^{(S_0)} >  = \vert S_{0} > \otimes (\otimes_{i=1}^b \vert \lambda^{-}_{i}(S_{0}) > )
\label{rgspin}
\end{eqnarray}
with the corresponding operators 
\begin{eqnarray}
\sigma_{R0}^z && = \vert \sigma_{R0}^z=+> <\sigma_{R0}^z=+ \vert
- \vert \sigma_{R0}^z=-> <\sigma_{R0}^z=- \vert
\nonumber \\
\sigma_{R0}^x  && =  \vert \sigma_{R0}^z=+> <\sigma_{R0}^z=- \vert
+ \vert \sigma_{R0}^z=-> <\sigma_{R0}^z=+ \vert
\label{sigmar}
\end{eqnarray}

\subsection{Projection rule for  $ \sigma^z_0 $  }

To evaluate $P_b \sigma^z_0 P_b$, we have to consider the action of 
the operator $\sigma^z_0 $ on each ground state 
\begin{eqnarray}
\sigma^z_0 \vert {GS}^{(S_0)}>
&& = \sigma^z_0 \vert S_{0} > \otimes (\otimes_{i=1}^b \vert \lambda^{-}_{i}(S_{0}) > )
\nonumber \\
&& = S_{0} \vert  S_{0} > \otimes (\otimes_{i=1}^b \vert \lambda^{-}_{i}(S_{0}) > )
\nonumber \\
&& =S_{0} \vert {GS}^{(S_0)}   >
\label{sznp1gs}
\end{eqnarray}
This yields the following trivial projection rule
in terms of the renormalized operator $\sigma_{R0}^z$ of Eq. \ref{sigmar}
\begin{eqnarray}
P_b \sigma^z_0 P_b =  \sigma^z_{R0}
\label{projsigmaz0}
\end{eqnarray}

\subsection{Projection rule for  $ \sigma^z_i $  }

To evaluate $P_b \sigma^z_{i} P_b$, we need to compute the action of 
the operator $\sigma^z_i $ on each ground state 
\begin{eqnarray}
\sigma^z_{i} \vert {GS}^{(S_0)}>
&& = \sigma^z_{i} \vert S_{0} > \otimes (\otimes_{j=1}^b \vert \lambda^{-}_{j}(S_{0}) > )
\nonumber \\
&& =  \vert S_{0} >  \otimes ( \sigma^z_{i} \vert \lambda^{-}_{i}(S_{0}) > ) \otimes (\otimes_{j \ne i}^b \vert \lambda^{-}_{j}(S_{0}) > )
\label{szRngs}
\end{eqnarray}

Using
\begin{eqnarray}
 < \lambda^{-}_{i}(S_{0}) \vert \sigma^z_{i} \vert \lambda^{-}_{i}(S_{0}) >
= \left( \frac{1-c^2_i(S_{0}) }{1+c^2_i(S_{0})}  \right)
= S_{0}     \frac{ J_{0,i}}{ \sqrt{ h_{i}^2 +J_{0,i}^2 }}
\label{prodscal}
\end{eqnarray}
one obtains the following projection rule
in terms of the renormalized operator $\sigma_{R0}^z$ of Eq. \ref{sigmar}
\begin{eqnarray}
P_b \sigma^z_{i}  P_b =
 \frac{ J_{0,i}}{ \sqrt{ h_{i}^2 +J_{0,i}^2 }}  \sigma^z_{R0}
\label{projsigmazi}
\end{eqnarray}

\subsection{ Projection rule for  $ \sigma^x_0 $  }

To compute $P_b \sigma^x_0 P_b$, we have to consider the action of 
the operator $\sigma^x_0 $ on each ground state 
\begin{eqnarray}
\sigma^x_0 \vert {GS}^{(S_0)}>
&& = \sigma^x_0 \vert S_{0} > \otimes (\otimes_{i=1}^b \vert \lambda^{-}_{i}(S_{0}) > )
\nonumber \\
&& = \vert - S_{0} > \otimes (\otimes_{i=1}^b \vert \lambda^{-}_{i}(S_{0}) > )
\label{sxnp1gs}
\end{eqnarray}
Using
\begin{eqnarray}
< {GS}^{( S_{0})} \vert\sigma^x_0 \vert {GS}^{(S_0)}> && = 0
\nonumber \\
< {GS}^{(- S_{0})} \vert\sigma^x_0 \vert {GS}^{(S_0)}>
&& = \prod_{i=1}^b <\lambda^{-}_{i}(-S_{0})  \vert \lambda^{-}_{i}(S_{0}) >
 = \prod_{i=1}^b \left( \frac{2 c_i(+)}{1+c_i^2(+)} \right)
= \prod_{i=1}^b \left(  \frac{ h_{i}}{ \sqrt{ h_{i}^2 +J_{0,i}^2 }}\right)
\label{sxnp1gsprod}
\end{eqnarray}
one obtains the following projection rule
in terms of the renormalized operator $\sigma_{R0}^x$ of Eq. \ref{sigmar}
\begin{eqnarray}
P_b \sigma^x_0 P_b = \left[
\prod_{i=1}^b \left(  \frac{ h_{i}}{ \sqrt{ h_{i}^2 +J_{0,i}^2 }} \right) \right] \sigma_{R0}^x
\label{projsigmax0}
\end{eqnarray}

\subsection{ Physical meaning}

The physical meaning of the procedure derived above
is thus very simple : the renormalized spin $\sigma_{R0}$ 
represents the 'master spin' $\sigma_{0}$ dressed by its b 'slave spins' $\sigma_{i}$.
For each slave $i=1,2,..,b$, we may consider the two limiting cases :

(a) if $J_{0,i} \gg h_i$, then the slave spin $\sigma_{i}$ is ferromagnetically locked to
its master $\sigma_{0}$ 
(Eq. \ref{projsigmazi} becomes $P_b \sigma^z_{i}  P_b \simeq  \sigma^z_{R0}$
to be compared with Eq. \ref{projsigmaz0}), so that the flipping of the master 
spin is affected by the small ratio $\frac{h_i}{J_{0,i}}$ (Eq. \ref{projsigmax0}).

(b) if $J_{0,i} \ll h_i$, then the slave spin $\sigma_{i}$ is mostly disordered and
only weakly polarized by its master $\sigma_{0}$ (Eq. \ref{projsigmazi} becomes $P_b \sigma^z_{i}  P_b \simeq \frac{J_{0,i}}{h_i}  \sigma^z_{R0}$), so that the flipping of the master spin is unchanged (Eq. \ref{projsigmax0}).

The projection rules above are thus compatible with the Strong Disorder RG rules in the two limits $J_{0,i} \gg h_i$ and  $J_{0,i} \ll h_i$ , 
but can also apply to cases where 
$J_{0,i} \sim h_i$. So they can be used to analyze both
 pure and random quantum Ising models, as recalled in the following section for $d=1$.

\section{ Reminder on the application in dimension $d=1$   }

\label{sec_d1}

The Hamiltonian of the quantum Ising chain reads
\begin{eqnarray}
H= - \sum_{i}  h (i) \sigma_{i}^x - \sum_{i}  J_{\vec x}(i) \sigma_{i}^z \sigma_{i+1}^z
\label{h1d}
\end{eqnarray}
In a block renormalization rule, one wishes to replace each block of two spins
$(\sigma_{2i-1};\sigma_{2i})$ by a single renormalized spin $\sigma_{R(2i)}$.
It is thus convenient to rewrite Eq. \ref{h1d} as
\begin{eqnarray}
H && =  \sum_{i} H_{i} 
 \\
 H_{i} && =  - h (2i) \sigma_{2i}^x -    h (2i-1) \sigma_{2i-1}^x
  -   \sigma_{2i-1}^z 
\left[  J_{\vec x}(2i-1) \sigma_{2i }^z + J_{\vec x}(2i-2) \sigma_{2i-2}^z   \right]
\label{h1di}
\end{eqnarray}

\subsection{ Block renormalization }

The idea of Fernandez-Pacheco \cite{pacheco}, 
written here for the random case \cite{nishiRandom}, is 
the following choice of the intra-block Hamiltonian
\begin{eqnarray}
H_{intra}^{(1)} && =  \sum_{(i)} H_{i}^{(1)} 
\nonumber \\
 H_{i}^{(1)}  && =  -    h (2i-1) \sigma_{2i-1}^x 
- J_{\vec x}(2i-1) \sigma_{2i-1}^z  \sigma_{2i }^z 
\label{h1intra1d}
\end{eqnarray}
Since $H_{i}^{(1)} $  has the form the Hamiltonian of Eq. \ref{hintra}
 analyzed in section \ref{sec_rgrule},
 the two spins $(\sigma_{2i};\sigma_{2i-1})$ can be renormalized
via a single renormalized spin $(\sigma_{R(2i)}^{R})$.
The application of the projection rules of Eqs \ref{projsigmaz0}, \ref{projsigmazi}
and \ref{projsigmax0} read for the present case
\begin{eqnarray}
P_{intra}^{(1)} \sigma^z_{2i} P_{intra}^{(1)} && =  \sigma^z_{R(2i)}
\nonumber \\
P_{intra}^{(1)} \sigma^z_{2i-1}  P_{intra}^{(1)} && =
 \frac{ J_{\vec x}(2i-1)}{ \sqrt{ h^2(2i-1)+J_{\vec x}^2(2i-1j) }}  \sigma^z_{R(2i)}
\nonumber \\
P_{intra}^{(1)} \sigma^x_{2i} P_{intra}^{(1)} && =
 \frac{h (2i-1)}{\sqrt{h^2(2i-1)+J_{\vec x}^2(2i-1) } } \sigma_{R(2i)}^x
\label{proj1dp1}
\end{eqnarray}

As a consequence, the projection of the remaining part of the Hamiltonian 
\begin{eqnarray}
H^R && = P_{intra}^{(1)} \left[ \sum_{(i)} (H_{i}-H_{i}^{(1)} )  \right] P_{intra}^{(1)} =  
\sum_{i} H_{i}^{R} 
\nonumber \\
H_{i}^{R} = &&   - h^R (2i)  \sigma_{R(2i)}^x
  - J_{2 \vec x}^R (2i-2)   \sigma^z_{R(2i-2)} \sigma^z_{R(2i)}
\label{h1dRfin}
\end{eqnarray}
has the same form as the initial Hamiltonian of Eq. \ref{h1d}, in terms of

(i)  the renormalized transverse fields on the remaining even sites
\begin{eqnarray}
h^R (2i) && =h(2i) 
 \frac{h (2i-1)}{\sqrt{h^2(2i-1)+J_{\vec x}^2(2i-1) } }
\label{rgh1d}
\end{eqnarray}

(ii) the renormalized couplings between the remaining even sites 
\begin{eqnarray}
 J^R_{2 \vec x} (2i-2) && = J_{ \vec x}(2i-2) 
\frac{J_{ \vec x} (2i-1)}{\sqrt{h^2(2i-1)+J^2_{ \vec x}(2i-1) } }
\label{rgj1d2x}
\end{eqnarray}

\subsection{ Application to the pure quantum Ising chain \cite{pacheco}}

If the initial parameters are $(h,J)$ on the whole chain,
 one obtains after one RG step the renormalized parameters (Eqs \ref{rgh1d} and \ref{rgj1d2x})
\begin{eqnarray}
h^R && = h  \frac{  h }{\sqrt{J^2 + h^2}}
\nonumber \\
J^R && = J  \frac{  J }{\sqrt{J^2 + h^2}}
\label{rgrulespur1d}
\end{eqnarray}
so that the ratio $K \equiv \frac{J}{h}$ evolves according to the simple rule
\begin{eqnarray}
K_R \equiv \frac{J_R}{h_R} =  K^2 \equiv \phi(K)
\label{d1pur}
\end{eqnarray}
The disordered attractive fixed point $K=0$ and the 
ferromagnetic attractive fixed point $K \to +\infty$ are separated by
the unstable fixed point $K_c=1$ characterized by the correlation length
exponent $\nu=1$ obtained by $2^{\frac{1}{\nu}}=\phi'(K_c)=2 K_c=2$.
The fact that both $K_c$ and $\nu$ are in agreement with the exact solution \cite{pfeuty} shows that the 
Fernandez-Pacheco choice \cite{pacheco} of the intra-block Hamiltonian of Eq. \ref{h1intra1d} is better than other choices  \cite{um,jullien,fradkin,hirsch}.

Since the quantum Ising chain represents the anisotropic limit of the two-dimensional classical Ising model and is the the same universality class, it is interesting to compare with all the real-space renormalization procedures concerning classical spin models, from the early Migdal-Kadanoff schemes 
to the more recent tensor networks formulations
(see the recent review \cite{rgreviewkadanoff} and references therein). For the two-dimensional Ising model, whenever the two directions are treated on the same footing, the various real-space RG procedures that have been proposed are able to produce very good approximations of the exponent $\nu$ (see for instance the Table I of the review \cite{rgreviewkadanoff}) but never yield exactly $\nu=1$, in contrast to the Fernandez-Pacheco quantum procedure described above. So it seems presently that the only way to obtain exactly $\nu=1$ for the 2D Ising model is by defining a renormalization procedure {\it for the transfer matrix} \cite{cardyrgtransfer}
in order to inherit the exactness of the exponent $\nu$ of the Fernandez-Pacheco quantum procedure.

\subsection{ Application to the disordered quantum Ising chain \cite{nishiRandom} }

In terms of the ratios
\begin{eqnarray}
K_{\vec x} (i) \equiv \frac{J_{x}(i)}{h_{i+1}} 
\label{ki}
\end{eqnarray}
the RG rules of Eq. \ref{rgh1d} and \ref{rgj1d2x} reads
\begin{eqnarray}
K^R_{2 \vec x}(2i-2) && \equiv \frac{J^R_{\vec x}(2i-2)}{h^R_{2i}} 
 = 
\frac{J_{ \vec x}(2i-2)  J_{ \vec x} (2i-1)} { h (2i-1) h(2i) }
  =K_{\vec x} (2i-2) K_{\vec x} (2i-1)  
\label{rgruleskr}
\end{eqnarray}
and thus corresponds to a simple addition in log-variables
\begin{eqnarray}
\ln K^R_{2 \vec x}(2i-2) =\ln K_{\vec x} (2i-2) + \ln K_{\vec x} (2i-1) 
\label{rgruleslogkr}
\end{eqnarray}

So after $N$ RG steps corresponding to a length $L=2^N$, 
the renormalized ratio $K^R(L)$ reads in terms of the initial variables
\begin{eqnarray}
\ln K^R_L = \sum_{i=1}^{L} \ln K_{\vec x} (i-1) =
 \sum_{i=1}^{L} \left[ \ln J_{\vec x}(i-1)-\ln h (i)   \right]
\label{inffp}
\end{eqnarray}
The Central Limit theorem yields the asymptotic behavior
\begin{eqnarray}
\ln K^R_L \opsimeq_{L \to + \infty} L \left[ \overline{ \ln J_{\vec x}(i-1) -\ln h (i) }  \right]
+ L^{1/2} \sqrt{ \left[ Var[\ln  J_{\vec x} ] + Var[\ln h ]   \right] } u
\label{inffpclt}
\end{eqnarray}
where $u$ is a Gaussian random variable.

The first term yields that the critical point corresponds to the condition
\begin{eqnarray}
 \overline{\ln J_{\vec x}(i-1) -\ln h(i)  }  =0
\label{criti1d}
\end{eqnarray}
and that the typical correlation length exponent is
\begin{eqnarray}
\nu_{typ}=1
\label{ntyp1d}
\end{eqnarray}

 Outside criticality, the competition between the first and the second term shows
 the finite-size correlation exponent is 
\begin{eqnarray}
\nu_{FS}=2
\label{nav1d}
\end{eqnarray}
At criticality where the first term vanishes, the second random
term of order $L^{1/2}$ corresponds to an Infinite Disorder Fixed Point of exponent
\begin{eqnarray}
\psi=\frac{1}{2}
\label{psi1d}
\end{eqnarray}

All these conclusions of Eqs \ref{criti1d}, \ref{ntyp1d}, \ref{nav1d}, \ref{psi1d}
obtained via the application of the Fernandez-Pacheco renormalization to
the random quantum Ising chain \cite{nishiRandom}, are
in agreement with
 the Fisher Strong Disorder renormalization exact results \cite{fisher}.
It is thus interesting to look for an appropriate generalization of 
the Fernandez-Pacheco renormalization in higher dimensions, and first of all in dimension $d=2$.

\section{ Block Renormalization Rules in dimension $d=2$   }

\label{sec_d2}

The initial quantum Ising Hamiltonian defined on the square lattice
of unit vectors $(\vec x,\vec y)$ reads
\begin{eqnarray}
H= - \sum_{(i,j)}  h (i,j) \sigma_{(i,j)}^x  
- \sum_{(i,j)} \sigma_{ (i,j)}^z \left[  J_{\vec x}(i,j) \sigma_{(i+1,j)}^z 
+  J_{\vec y}(i,j) \sigma_{(i,j)}^z    \right]
\label{h2d}
\end{eqnarray}
Various generalizations of the one-dimensional Fernandez-Pacheco renormalization
procedure have been already proposed in dimension $d=2$, both for the pure case
 \cite{epele,nishiPur,kubica} and for the disordered case \cite{nishiRandom},
with the drawbacks recalled in the Introduction.
In this section, we thus introduce another procedure where the two directions are 
considered on the same footing.

We wish to define a block renormalization rule, where each block of four spins
$(\sigma_{2i,2j};\sigma_{2i-1,2j};\sigma_{2i,2j-1};\sigma_{2i-1,2j-1}$ will be replaced
by a single renormalized spin $(\sigma_{2i,2j}^{RR})$, after two elementary 
renormalization steps.
It is convenient to start by rewriting Eq. \ref{h2d} as
\begin{eqnarray}
 H  && =  \sum_{(i,j)} H_{i,j} 
\label{h2dbox}
 \\
 H_{i,j}  && =  - h (2i,2j) \sigma_{(2i,2j)}^x -    h (2i-1,2j-1) \sigma_{(2i-1,2j-1)}^x
 -  h (2i-1,2j) \sigma_{(2i-1,2j)}^x -  h (2i,2j-1) \sigma_{(2i,2j-1)}^x
\nonumber \\
&&  -   \sigma_{(2i-1,2j)}^z 
[  J_{\vec x}(2i-1,2j) \sigma_{(2i,2j) }^z +  J_{\vec y}(2i-1,2j) \sigma_{2i-1,2j+1}^z 
 \nonumber \\ &&
\ \ \ \ \ \ \ \  \ \ \ \  \ \  + J_{\vec x}(2i-2,2j) \sigma_{(2i-2,2j)}^z + J_{\vec y}(2i-1,2j-1) \sigma_{(2i-1,2j-1)}^z  ]
\nonumber \\
&&  -   \sigma_{(2i,2j-1)}^z 
[  J_{\vec x}(2i,2j-1) \sigma_{(2i+1,2j-1) }^z +  J_{\vec y}(2i,2j-1) \sigma_{2i,2j}^z 
 \nonumber \\ &&
\ \ \ \ \ \ \ \  \ \ \ \  \ \   + J_{\vec x}(2i-1,2j-1)\sigma_{(2i-1,2j-1)}^z + J_{\vec y}(2i,2j-2) \sigma_{(2i,2j-2)}^z   ]
\nonumber
\end{eqnarray}

\subsection{ First renormalization step }

For the first renormalization step, we choose the following intra-block Hamiltonian
\begin{eqnarray}
H_{intra}^{(1)} && =  \sum_{(i,j)} H_{i,j}^{(1)} 
\nonumber \\
H_{i,j}^{(1)} && \equiv 
- h (2i-1,2j) \sigma_{2i-1,2j}^x -  J_{\vec x} (2i-1,2j)\sigma_{2i-1,2j}^z \sigma_{2i,2j}^z 
\nonumber \\
&& - h (2i,2j-1) \sigma_{2i,2j-1}^x - J_{\vec y}(2i,2j-1)  \sigma_{2i,2j-1}^z \sigma_{2i,2j}^z 
\label{h1intra2d}
\end{eqnarray}
Since $H_{i,j}^{(1)} $  has the form the Hamiltonian of Eq. \ref{hintra} analyzed in section \ref{sec_rgrule},
 the three spins $(\sigma_{2i,2j};\sigma_{2i-1,2j};\sigma_{2i,2j-1})$ can be renormalized
via a single renormalized spin $(\sigma_{2i,2j}^{R})$, whereas the spin $\sigma_{2i-1,2j-1} $
that is not involved in $H_{i,j}^{(1)} $ remains unchanged.
The application of the projection rules of Eqs \ref{projsigmaz0}, \ref{projsigmazi}
and \ref{projsigmax0} read for the present case
\begin{eqnarray}
P_{intra}^{(1)} \sigma^z_{2i,2j} P_{intra}^{(1)} && =  \sigma^z_{R(2i,2j)}
\nonumber \\
P_{intra}^{(1)} \sigma^z_{2i-1,2j}  P_{intra}^{(1)} && =
 \frac{ J_x(2i-1,2j)}{ \sqrt{ h^2(2i-1,2j)+J^2_{\vec x}(2i-1,2j) }}  \sigma^z_{R(2i,2j)}
\nonumber \\
P_{intra}^{(1)} \sigma^z_{2i,2j-1}  P_{intra}^{(1)} && =
 \frac{ J_y(2i,2j-1)}{ \sqrt{ h^2(2i,2j-1)+J^2_{\vec y}(2i,2j-1) }}  \sigma^z_{R(2i,2j)}
\nonumber \\
P_{intra}^{(1)} \sigma^x_{2i,2j} P_{intra}^{(1)} && =
 \frac{h (2i-1,2j)}{\sqrt{h^2(2i-1,2j)+J^2_{\vec x}(2i-1,2j) } }
\frac{h (2i,2j-1)}{\sqrt{h^2(2i,2j-1)+J^2_{\vec y}(2i,2j-1) } } \sigma_{R(2i,2j)}^x
\label{proj2dp1}
\end{eqnarray}

As a consequence, the projection of the remaining part of the Hamiltonian reads
\begin{eqnarray}
H^R && = P_{intra}^{(1)} \left[ \sum_{(i,j)} (H_{i,j}-H_{i,j}^{(1)} )  \right] P_{intra}^{(1)} =  
\sum_{(i,j)} H_{i,j}^{R} 
\nonumber \\
H_{i,j}^{R} = && -  h (2i-1,2j-1) \sigma_{(2i-1,2j-1)}^x   - h^R (2i,2j)  \sigma_{R(2i,2j)}^x
\nonumber \\
&&  - J^R_{2 \vec x} (2i-2,2j)   \sigma^z_{R(2i-2,2j)} \sigma^z_{R(2i,2j)}
 -  J^R_{2\vec y}(2i,2j-2) \sigma^z_{R(2i,2j-2)}  \sigma^z_{R(2i,2j)}
\nonumber \\
&&  - 
J^R_{ \vec x- \vec y} (2i,2j)
  \sigma^z_{R(2i,2j)}  \sigma_{(2i+1,2j-1) }^z  - J^R_{ -\vec x+ \vec y} (2i,2j)  
 \sigma^z_{R(2i,2j)} \sigma_{2i-1,2j+1}^z    
\nonumber \\
&& -  J^R_{ \vec x+ \vec y} (2i-1,2j-1) \sigma_{(2i-1,2j-1)}^z  \sigma^z_{R(2i,2j)}
\label{h2dRfin}
\end{eqnarray}
in terms of 

(i)  the renormalized transverse fields of even-even sites 
\begin{eqnarray}
h^R (2i,2j) && =h(2i,2j) 
\frac{h (2i-1,2j)}{\sqrt{h^2(2i-1,2j)+J^2_{\vec x}(2i-1,2j) } }
\frac{h (2i,2j-1)}{\sqrt{h^2(2i,2j-1)+J^2_{\vec y}(2i,2j-1) } }
\label{rgh2d}
\end{eqnarray}

(ii) the renormalized couplings along the horizontal directions at distance two
\begin{eqnarray}
J^R_{2 \vec x} (2i-2,2j) && = J_{ \vec x}(2i-2,2j) 
\frac{J_{ \vec x} (2i-1,2j)}{\sqrt{h^2(2i-1,2j)+J^2_{ \vec x}(2i-1,2j) } }
\label{rgj2d2x}
\end{eqnarray}

(iii) the renormalized couplings along the vertical directions at distance two
\begin{eqnarray}
J^R_{2 \vec y} (2i,2j-2) && = J_{ \vec y}(2i,2j-2) 
\frac{J_{ \vec y} (2i,2j-1)}{\sqrt{h^2(2i,2j-1)+J^2_{ \vec y}(2i,2j-1) } }
\label{rgj2d2y}
\end{eqnarray}

(iv) the renormalized couplings along the diagonal directions $(\vec x- \vec y)$ and $(-\vec x+ \vec y)$
\begin{eqnarray}
J^R_{ \vec x- \vec y} (2i,2j) && = J_{ \vec x}(2i,2j-1) 
\frac{J_{ \vec y} (2i,2j-1)}{\sqrt{h^2(2i,2j-1)+J^2_{ \vec y}(2i,2j-1) } }
\nonumber \\
J^R_{ -\vec x+ \vec y} (2i,2j) && = J_{ \vec y}(2i-1,2j) 
\frac{J_{ \vec x} (2i-1,2j)}{\sqrt{h^2(2i-1,2j)+J^2_{ \vec x}(2i-1,2j) } }
\label{rgj2ddiagexter}
\end{eqnarray}

(v) the renormalized couplings along the diagonal direction $(\vec x+ \vec y)$ within each block
\begin{eqnarray}
J^R_{ \vec x+ \vec y} (2i-1,2j-1) && = J_{ \vec x}(2i-1,2j-1) 
\frac{J_{ \vec y} (2i,2j-1)}{\sqrt{h^2(2i,2j-1)+J^2_{ \vec y}(2i,2j-1) } }
 \nonumber \\ &&
+  J_{ \vec y}(2i-1,2j-1) 
\frac{J_{ \vec x} (2i-1,2j)}{\sqrt{h^2(2i-1,2j)+J^2_{ \vec x}(2i-1,2j) } }
\label{rgj2ddiaginter}
\end{eqnarray}

\subsection{ Second renormalization step }

For the Hamiltonian $H^R$ of Eq. \ref{h2dRfin}, we choose the following intra-block Hamiltonian
\begin{eqnarray}
H_{i,j}^{(2)} \equiv 
- h (2i-1,2j-1) \sigma_{2i-1,2j-1}^x 
 - J^R_{\vec x+\vec y} (2i-1,2j-1) \sigma_{2i-1,2j-1}^z \sigma_{R(2i,2j)}^z 
\label{h2intra2d}
\end{eqnarray}
It has the form the Hamiltonian of Eq. \ref{hintra} analyzed in section \ref{sec_rgrule},
so that the two spins $(\sigma^R_{2i,2j};\sigma_{2i-1,2j-1})$ can be replaced by a single renormalized spin $(\sigma_{2i,2j}^{RR})$.
The application of the projection rules of Eqs \ref{projsigmaz0}, \ref{projsigmazi}
and \ref{projsigmax0} read for the present case
\begin{eqnarray}
P_{intra}^{(2)} \sigma^z_{R(2i,2j)} P_{intra}^{(2)} && =  \sigma^z_{RR(2i,2j)}
\nonumber \\
P_{intra}^{(2)} \sigma^z_{2i-1,2j-1}  P_{intra}^{(2)} && =
 \frac{ J^R_{\vec x+\vec y} (2i-1,2j-1) }
{ \sqrt{ h^2(2i-1,2j-1)+[J^R_{\vec x+\vec y} (2i-1,2j-1)]^2 }}  \sigma^z_{RR(2i,2j)}
\nonumber \\
P_{intra}^{(2)} \sigma^x_{R(2i,2j)} P_{intra}^{(2)} && =
 \frac{h (2i-1,2j-1)}{\sqrt{h^2(2i-1,2j-1)+[J^R_{\vec x+\vec y} (2i-1,2j-1)]^2 } } \sigma_{RR(2i,2j)}^x
\label{proj2dp2}
\end{eqnarray}

As a consequence, the projection of the remaining part of the Hamiltonian reads
\begin{eqnarray}
H^{RR} && = P_{intra}^{(2)} \left[ \sum_{(i,j)} (H_{i,j}^R-H_{i,j}^{(2)} )  \right] P_{intra}^{(2)} =  
\sum_{(i,j)} H_{i,j}^{RR} 
\nonumber \\
H_{i,j}^{RR} = &&    - h^{RR} (2i,2j)   \sigma_{RR(2i,2j)}^x
 - J^R_{2 \vec x} (2i-2,2j)   \sigma^z_{RR(2i-2,2j)} \sigma^z_{RR(2i,2j)}
 -  J^R_{2\vec y}(2i,2j-2) \sigma^z_{RR(2i,2j-2)}  \sigma^z_{RR(2i,2j)}
\label{h2dRRfin}
\end{eqnarray}
i.e. it has the same form as the initial Hamiltonian on the square lattice,
in terms of

(i)  the renormalized transverse fields 
\begin{eqnarray}
h^{RR} (2i,2j) && =h^R(2i,2j) 
\frac{h (2i-1,2j-1)}{\sqrt{[h(2i-1,2j-1)]^2+[J^R_{\vec x+\vec y}(2i-1,2j-1)]^2 } }
\label{rghRR}
\end{eqnarray}

(ii) the renormalized couplings along the horizontal directions at distance two
\begin{eqnarray}
J^{RR}_{2 \vec x} (2i-2,2j) && = J^R_{2 \vec x}(2i-2,2j)
+ J^R_{ \vec x- \vec y} (2i-2,2j)
\frac{J^R_{ \vec x+ \vec y} (2i-1,2j-1)}{\sqrt{[h(2i-1,2j-1)]^2
+[J^R_{ \vec x+ \vec y}(2i-1,2j-1)]^2 } }
\label{rgj2ddRRh}
\end{eqnarray}

(ii) the renormalized couplings along the vertical directions at distance two
\begin{eqnarray}
J^{RR}_{2 \vec y} (2i,2j-2) && = J^R_{2 \vec y}(2i,2j-2)
+ J^R_{ -\vec x+ \vec y} (2i,2j-2)  
    \frac{ J^R_{\vec x+\vec y} (2i-1,2j-1) }
{ \sqrt{ h^2(2i-1,2j-1)+[J^R_{\vec x+\vec y} (2i-1,2j-1)]^2 }} 
\label{rgj2ddRRv}
\end{eqnarray}

 In the following sections, we discuss the application of these renormalization rules to the pure case, to the random ferromagnetic case, and to the spin-glass case.

\section{ Application to the pure two-dimensional quantum Ising model }

\label{sec_d2pure}

If we start from the pure model of parameters $(h,J)$,
the renormalization rules of Eqs \ref{rghRR} \ref{rgj2ddRRh} \ref{rgj2ddRRv} reduce to
\begin{eqnarray}
h^{RR} && = h \frac{h^2 }{ h^2+J^2 } 
\frac{h}{\sqrt{h^2+[ 2 J  \frac{J}{\sqrt{h^2+J^2 } }]^2 } } =  \frac{h^4 }{\sqrt{h^2+J^2} \sqrt{h^2(h^2+J^2)+4 J^4  } }
\nonumber \\ 
J^{RR}  && =   \frac{J^2}{\sqrt{h^2+J^2 }}
\left[ 1+  \frac{   2 J^2   }{\sqrt{h^2(h^2+J^2)+  4J^4 }}\right]
\label{purrg2d}
\end{eqnarray}
In terms of the ratio $K \equiv \frac{J}{h}$, the renormalization rule
reads
\begin{eqnarray}
K^{RR} \equiv \frac{J^{RR} }{h^{RR} } = K^2
\left[\sqrt{ 1+K^2+4 K^4 } +  2 K^2  \right] \equiv \phi(K)
\label{xevol2d}
\end{eqnarray}

We are now interested into the critical point satisfying the fixed point equation
$K_c=\phi(K_c)$ (between the ferromagnetic fixed point $K_*=+\infty$
and the disordered fixed point $K_*=0$) : eliminating the square-root yields
the fourth degree equation
\begin{eqnarray}
   K_c^4+ 4 K_c^3 + K_c^2 -1 =0
\label{eqkc2d}
\end{eqnarray}
The only positive root reads (the other three roots can be disregarded since one is negative, and the two others are imaginary) 
\begin{eqnarray}
u && \equiv (18 \sqrt{103}-179 )^{1/3}
\nonumber \\ 
v && \equiv \sqrt{10+u-\frac{11}{u}}
\nonumber \\
K_c && = \frac{1}{6} \sqrt{90+ \frac{108 \sqrt{3}}{v} - 3 v^2}   - \frac{v}{2 \sqrt{3}}-1
\simeq 0.538752
\label{reskc2d}
\end{eqnarray}
The correlation length exponent $\nu$ can be obtained from
\begin{eqnarray}
2^{\frac{1}{\nu}}  =\phi'(K_c) = \frac{K_c^2 ( 2 +8 K_c +3 K_c^2) }{ 1- 2 K_c^3}
\label{nu2d}
\end{eqnarray}
and the corresponding numerical value
\begin{eqnarray}
\nu \simeq 0.624758..
\label{nu2dnume}
\end{eqnarray}
is very close to the numerical estimate $\nu \simeq 0.63$ for the 3D classical Ising model. 

The two-dimensional procedure that we have proposed is thus much simpler
that the Miyazaki-Nishimori-Ortiz procedure that needs re-symmetrization between the two directions \cite{nishiPur} and yields a very good approximation for the
correlation length exponent $\nu$, better than other real-space renormalization
procedures \cite{penson,hirsch2d,mattis}.
The extension to the pure model in $d=3$ is described in the Appendix A,
and we now turn to the random models in dimension $d=2$.

\section{ Application to the random ferromagnetic two-dimensional quantum Ising model }

\label{sec_d2randomferro}

\subsection{ Numerical details }

In order to compare with the Strong Disorder renormalization numerical results
 \cite{kovacs2d,kovacsreview},
we have adopted the standard choice of a flat distribution of couplings between $J=0$ and $J=1$
\begin{eqnarray}
P(J) = \theta(0 \leq J \leq 1)
\label{pJflat}
\end{eqnarray}
and a flat distribution of transverse fields between $h=0$ and $h_b$
\begin{eqnarray}
Q(h) = \frac{1}{h_b} \theta(0 \leq h \leq h_b)
\label{phflat}
\end{eqnarray}
so that the control parameter of the zero-temperature transition is
\begin{eqnarray}
\theta \equiv \ln h_b
\label{control}
\end{eqnarray}

We have applied numerically the renormalization rules derived above
to $n_s=25 000 $ disordered two-dimensional samples of linear size 
\begin{eqnarray}
L_s=2^{12}=4096
\label{sizes}
\end{eqnarray}
 (containing $L_s^2=2^{24} $ spins) with periodic boundary conditions.
The renormalization procedure is stopped at the scale $L=2^{11}$, where there remains only four sites and eight links in each sample. As a consequence for this largest length $L=2^{11}$, the statistics is over $4 n_s=10^5$ random fields and over $8 n_s = 2 \times 10^5$ random couplings.

At each renormalization step corresponding
 to the lengths $L=2^n$ with $0 \leq n \leq 11$
we have analyzed the statistical properties of the renormalized transverse fields and
of the renormalized couplings.
More precisely, we have measured the RG flows of the typical values defined by
\begin{eqnarray}
\ln h_L^{typ} && \equiv \overline{ \ln h_L } 
\nonumber \\
\ln J_L^{typ} && \equiv \overline{ \ln J_L }  
\label{deftyp}
\end{eqnarray}
and of the widths of the probability distributions 
\begin{eqnarray}
\Delta_{\ln h_L} && \equiv  \sqrt{ \overline{ (\ln h_L)^2 } - (\overline{ \ln h_L })^2 }
\nonumber \\
\Delta_{\ln J_L} && \equiv  \sqrt{ \overline{ (\ln J_L)^2 } - (\overline{ \ln J_L })^2 } 
\label{defwidth}
\end{eqnarray}
as a function of the length $L$
for 28 values of the control parameter $\theta$ of Eq. \ref{control}.

The linear size $L_s=4096$ and the statistics over  $n_s=25 000 $
are thus of the same order of those used in recent Strong Disorder Renormalization studies in $d=2$ \cite{kovacs2d,kovacsreview}, but of course the implementation is much simpler here since the spatial structure remains a square lattice upon RG
 instead of an evolving non-trivial topology. 
Another difference is that  we analyze the statistics over samples at fixed control parameter $\theta$ and fixed size $L$, whereas Strong Disorder Renormalization studies of Ref. \cite{kovacs2d,kovacsreview} are based on the determination of
the pseudo-critical parameter for each sample.

\subsection{ RG flow of the renormalized transverse fields }

\begin{figure}[htbp]
 \includegraphics[height=6cm]{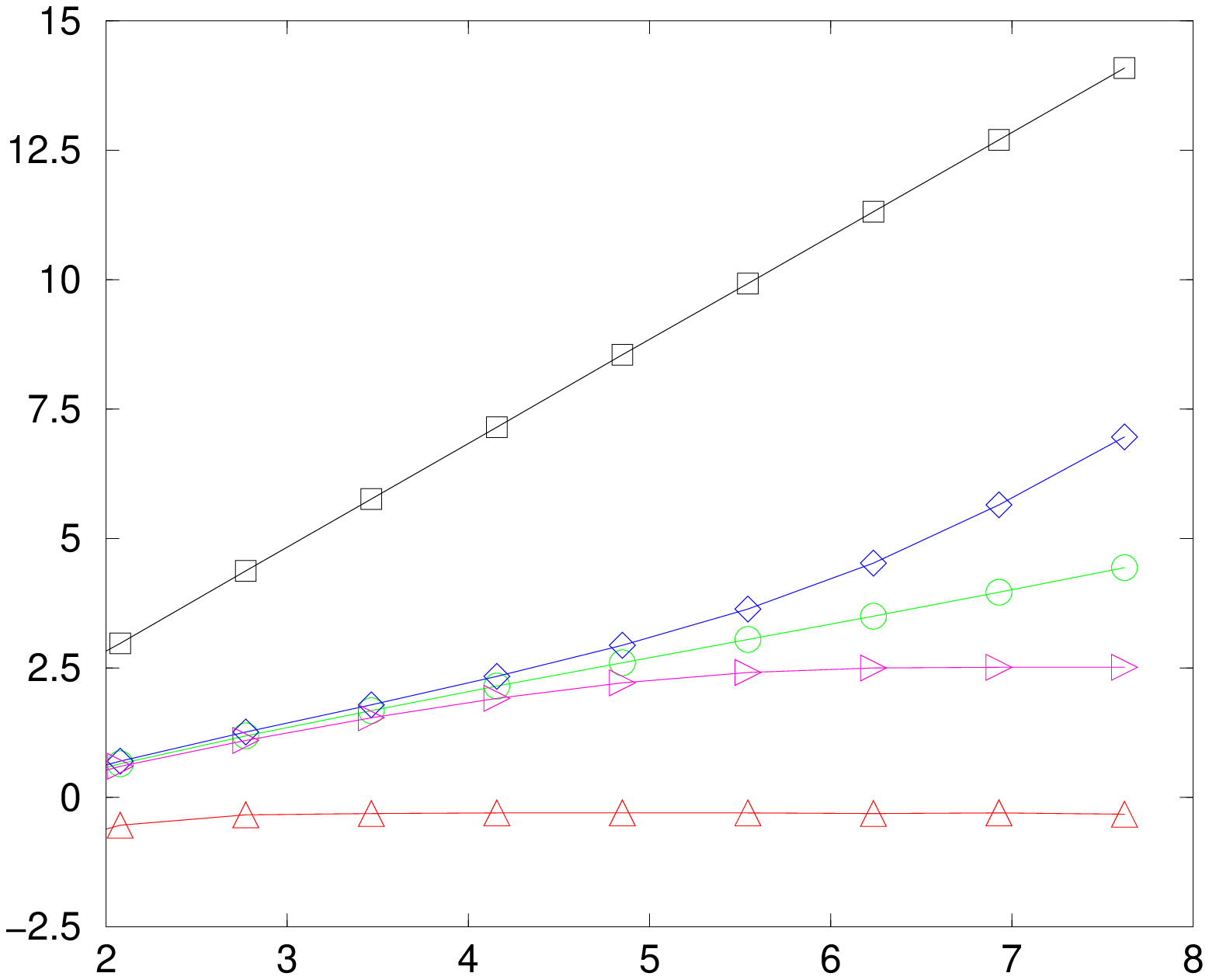}
\hspace{1cm}
 \includegraphics[height=6cm]{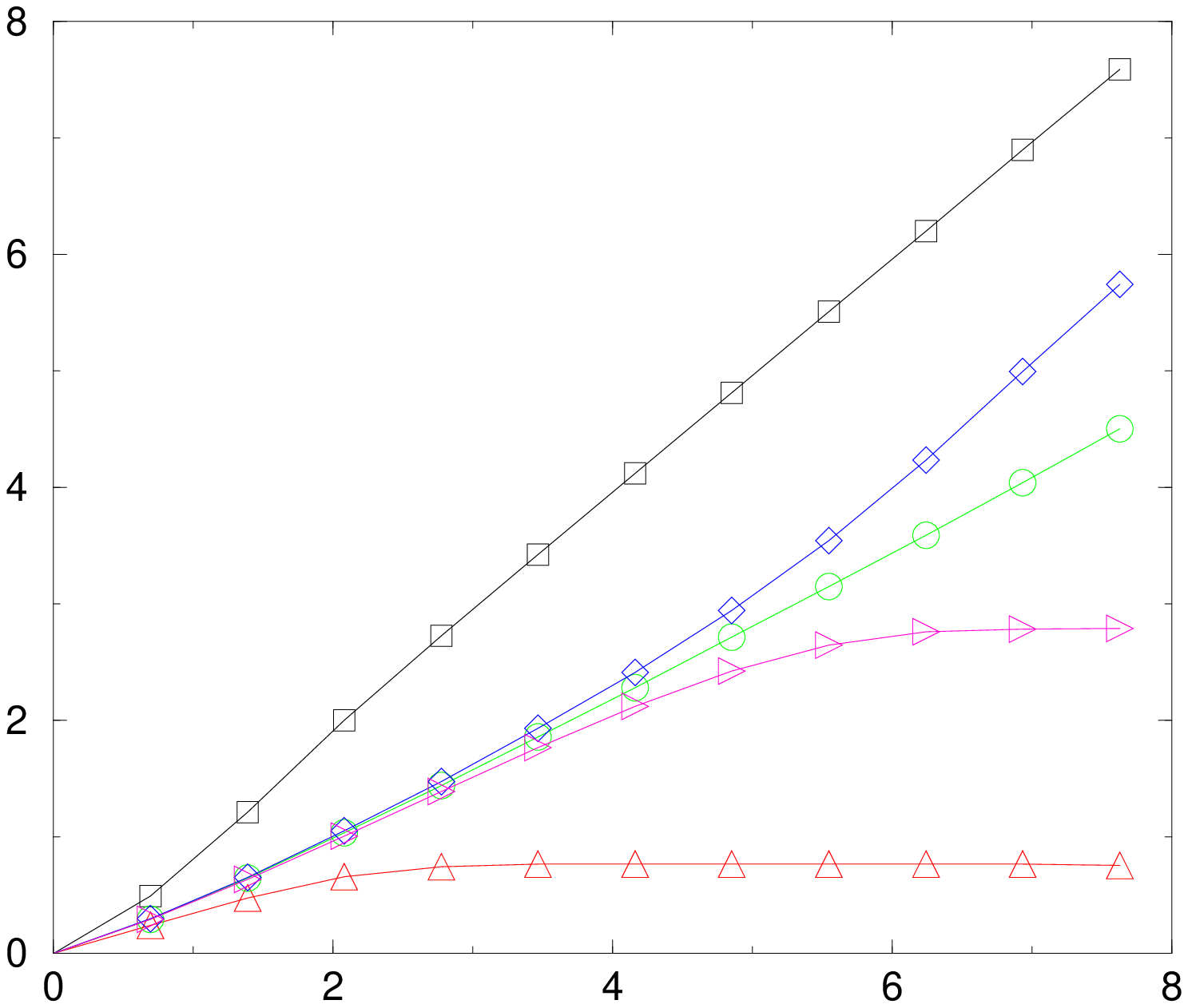}
\caption{ RG flow of the logarithm of the transverse fields in log-log scale \\
(a)  $Y_a= \ln (- \ln h_L^{typ}) \equiv \ln (- \overline{\ln h_L} )$ as a function of $X=\ln L$ : \\
(i) Ordered phase $\theta<\theta_c$ :   the asymptotic slope is $d=2$, as shown here for $\theta=0.5$ (squares) and $\theta=1.245$ (diamond) \\
(ii) Disordered phase $\theta>\theta_c$ :   the asymptotic slope is $0$, as shown here for $\theta=1.5$ (triangles up) and $\theta=1.27$ (triangles right) \\
(iii) Critical point $\theta_c =1.256$ (circles) : the asymptotic slope is
  $\psi \simeq 0.65 $. \\
(b)   $Y_b = \ln ( \Delta_{\ln h_L}) \equiv \ln (\sqrt{ \overline{(\ln h_L)^2} -(\overline{\ln h_L})^2 })$ as a function of $X= \ln L$ : \\
(i) Ordered phase $\theta<\theta_c$ :   the asymptotic slope is $1$, as shown here for $\theta=0.5$ (squares) and $\theta=1.245$ (diamond) \\
(ii) Disordered phase $\theta>\theta_c$ :   the asymptotic slope is $0$, as shown here for $\theta=1.5$ (triangles up) and $\theta=1.27$ (triangles right) \\
(iii) Critical point $\theta_c =1.256$ (circles) : the asymptotic slope is
  $\psi \simeq 0.65 $.
  }
\label{figflowh}
\end{figure}

On Fig. \ref{figflowh}, we show in log-log scale the RG flows of 
the typical renormalized transverse field $h_L^{typ}$ of Eq. \ref{deftyp}
\begin{eqnarray}
\ln h_L^{typ} \vert_{\theta<\theta_c}  && \oppropto_{L \to +\infty} -L^2     \nonumber \\
\ln h_L^{typ} \vert_{\theta=\theta_c}  && \oppropto_{L \to +\infty} - L^{\psi} \ \ {\rm with } \ \ \psi \simeq 0.65  \nonumber \\
\ln h_L^{typ} \vert_{\theta>\theta_c} && \oppropto_{L \to +\infty} Cst 
\label{htypflow}
\end{eqnarray}
and of the width $\Delta_{\ln h_L} $ of Eq. \ref{defwidth}
\begin{eqnarray}
\Delta_{\ln h_L} \vert_{\theta<\theta_c}  && \oppropto_{L \to +\infty} L     \nonumber \\
\Delta_{\ln h_L} \vert_{\theta=\theta_c}  && \oppropto_{L \to +\infty}  L^{\psi} \ \ {\rm with } \ \ \psi \simeq 0.65  \nonumber \\
\Delta_{\ln h_L} \vert_{\theta>\theta_c} && \oppropto_{L \to +\infty} Cst 
\label{hwidthflow}
\end{eqnarray}

At the critical point $\theta_c $, 
the RG flows of the typical value and of the width 
display the same activated scaling of an Infinite Disorder Fixed Point
\begin{eqnarray}
\ln h_L\vert_{\theta=\theta_c} && = -  L^{\psi} v_c
\label{hLcriti}
\end{eqnarray}
where $v_c$ is some $O(1)$ random variable.
The values obtained here for the location
 of the critical point $\theta_c \simeq 1.256$ 
and the activated exponent $\psi \simeq 0.65 $ turn out to be different from
the estimations $\theta_c^{SD} \simeq 1.678$ and $\psi^{SD} \simeq 0.48$
obtained via the Strong Disorder Renormalization
 (see \cite{kovacs2d,kovacsreview} and references therein) : 
the origin of these differences is not clear to us.

In the ordered phase, 
the logarithm of the typical renormalized transverse field grows extensively
with respect to the volume $L^d$ (with $d=2$ here)
\begin{eqnarray}
\ln h_L^{typ} \oppropto_{L \to \infty}   - \left( \frac{L}{\xi_h} \right)^d 
\label{hLorder}
\end{eqnarray}
where the length scale $\xi_h$ represents the characteristic size of finite disordered clusters
within this ordered phase. From our numerical data concerning the ordered phase, 
 the asymptotic behavior of Eq. \ref{hLorder} allows to measure $\xi_h$ and its divergence
near criticality 
\begin{eqnarray}
\xi_h \oppropto_{\theta \to \theta_c} (\theta_c-\theta)^{- \nu_h} \ \ \ {\rm with } \ \ \ \nu_h \simeq 0.84
\label{nuh}
\end{eqnarray}

In the disordered phase, the asymptotic typical value $h_{\infty}^{typ} $
 diverges with an essential singularity as a function of the control parameter
\begin{eqnarray}
\ln h_{\infty}^{typ} \equiv \overline{\ln h_{\infty}} \propto   - (\theta-\theta_c)^{-\kappa} 
\ \ \ {\rm  with } \ \ \ \kappa \simeq 0.82
\label{defkappa}
\end{eqnarray}

\begin{figure}[htbp]
 \includegraphics[height=6cm]{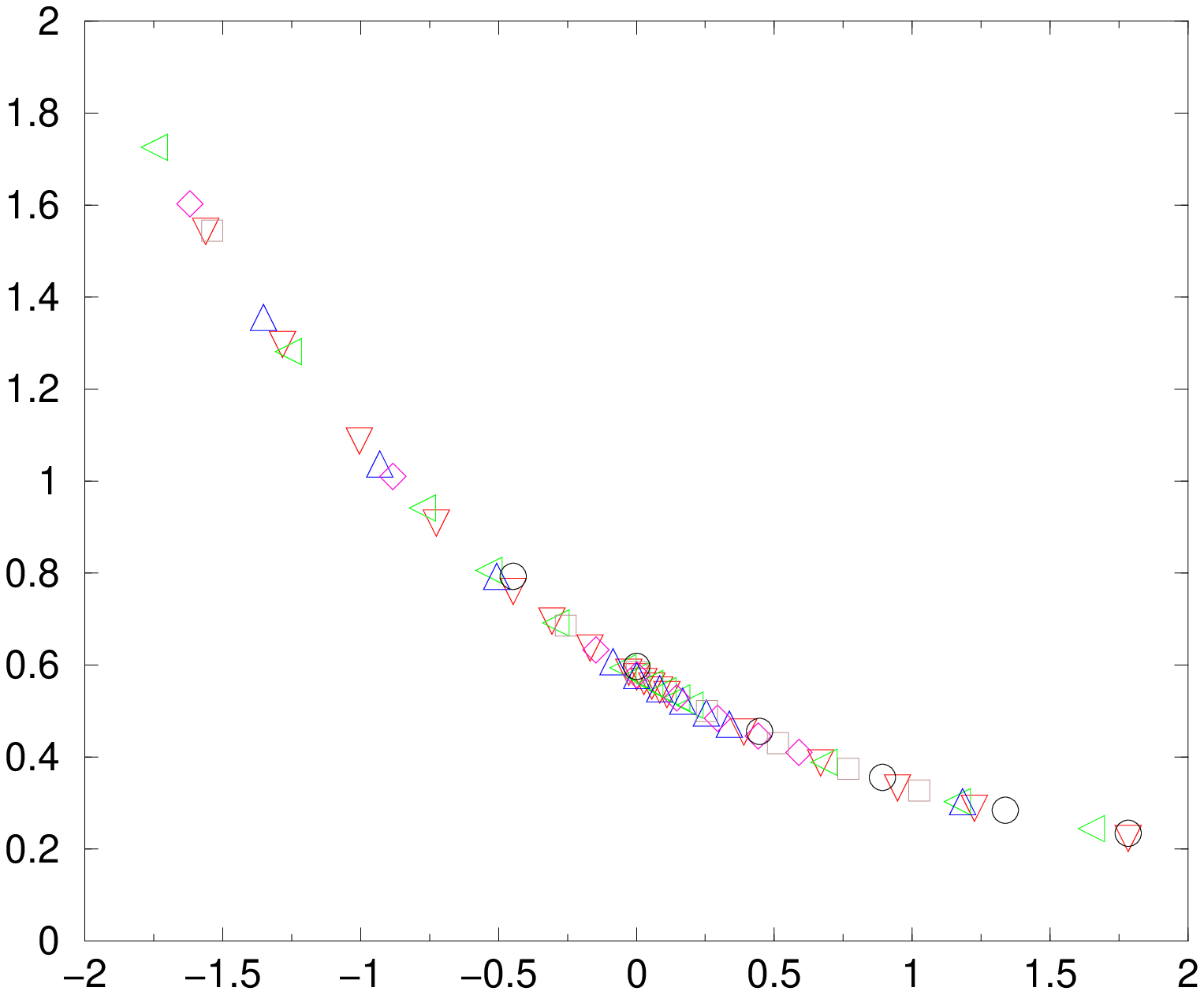}
\hspace{1cm}
 \includegraphics[height=6cm]{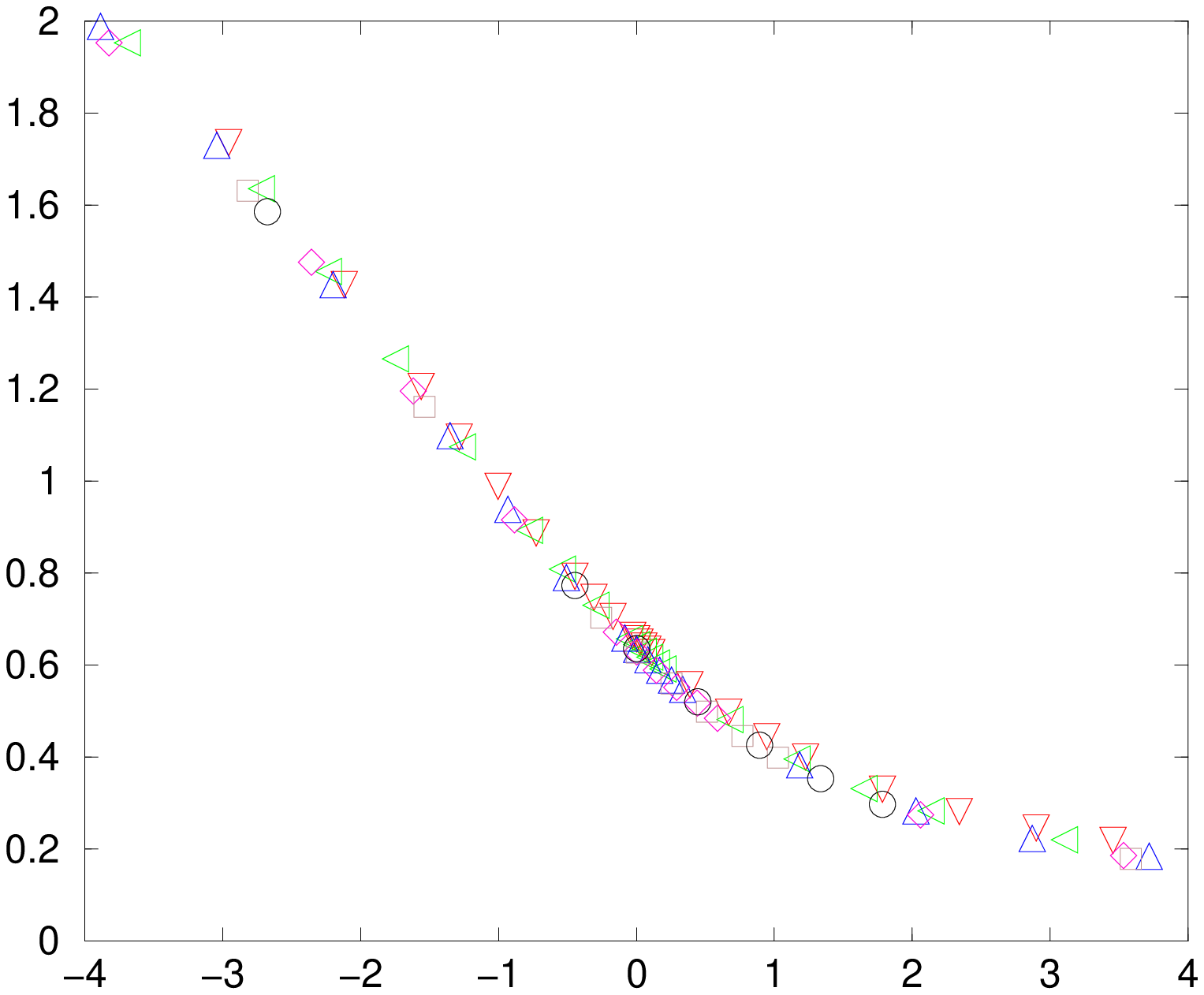}
\caption{ Finite-size scaling plots of the numerical data of Fig. \ref{figflowh} corresponding to the sizes $2^6 \leq L \leq 2^{11}$
with the values $\theta_c =1.256$ for the critical point and $\nu_{FS}=1.25$
for the finite-size correlation length exponent :
 \\
(a)  $Y_a= \left(- \frac{ \ln h_L^{typ}}{L^{\psi}}\right)$  as a function of $X=(\theta-\theta_c) L^{\frac{1}{\nu_{FS}}}$.\\
(b)  $ Y_b= \left( \frac{\Delta_{\ln h_L} }{L^{\psi}}\right)$  as a function of $X=(\theta-\theta_c) L^{\frac{1}{\nu_{FS}}}$.
  }
\label{figfss}
\end{figure}

In the critical region, the finite-size scaling form is governed by some finite-size scaling
correlation length exponent $\nu_{FS}$
\begin{eqnarray}
\ln h_L^{typ} \equiv \overline{ \ln h_L }  = -  L^{\psi} F_h\left( L^{\frac{1}{\nu_{FS}}} ( \theta-\theta_c ) \right)
\label{fsshtyp}
\end{eqnarray}
The matching with the behavior in the ordered phase (Eqs \ref{hLorder} and \ref{nuh})
and in the disordered phase (Eq. \ref{defkappa})
yields the relations
\begin{eqnarray}
\nu_{h} = \left( 1- \frac{\psi}{d}\right) \nu_{FS}
\label{nuhnufs}
\end{eqnarray}
and
\begin{eqnarray}
\kappa = \psi \nu_{FS}
\label{kappafss}
\end{eqnarray}
The previous numerical measures of $\psi$, $\nu_h$ and $\kappa$ yield the estimate
\begin{eqnarray}
 \nu_{FS} \simeq 1.25
\label{nufss}
\end{eqnarray}
As shown on Fig. \ref{figfss}, this value of $\nu_{FS}$
 gives satisfactory
 finite-size scaling plots of the numerical data of Fig. \ref{figflowh}.

This value for $\nu_{FS}$ (Eq. \ref{nufss}) agrees 
with the estimations obtained via the Strong Disorder Renormalization
 (see \cite{kovacs2d,kovacsreview} and references therein),
and with the asymmetric block renormalization of Ref \cite{nishiRandom}.

\subsection{ RG flow of the renormalized couplings }

\begin{figure}[htbp]
 \includegraphics[height=6cm]{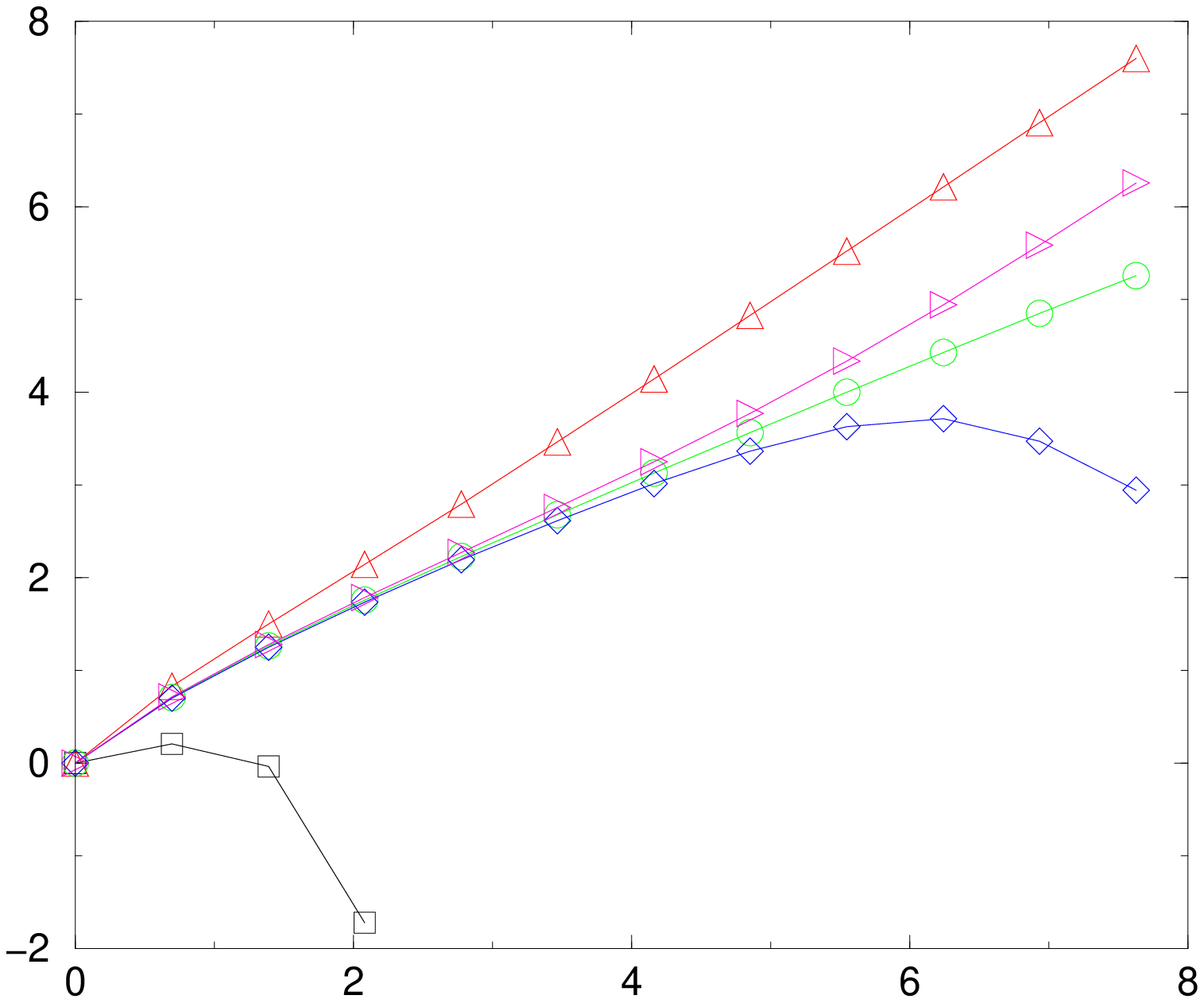}
\hspace{1cm}
 \includegraphics[height=6cm]{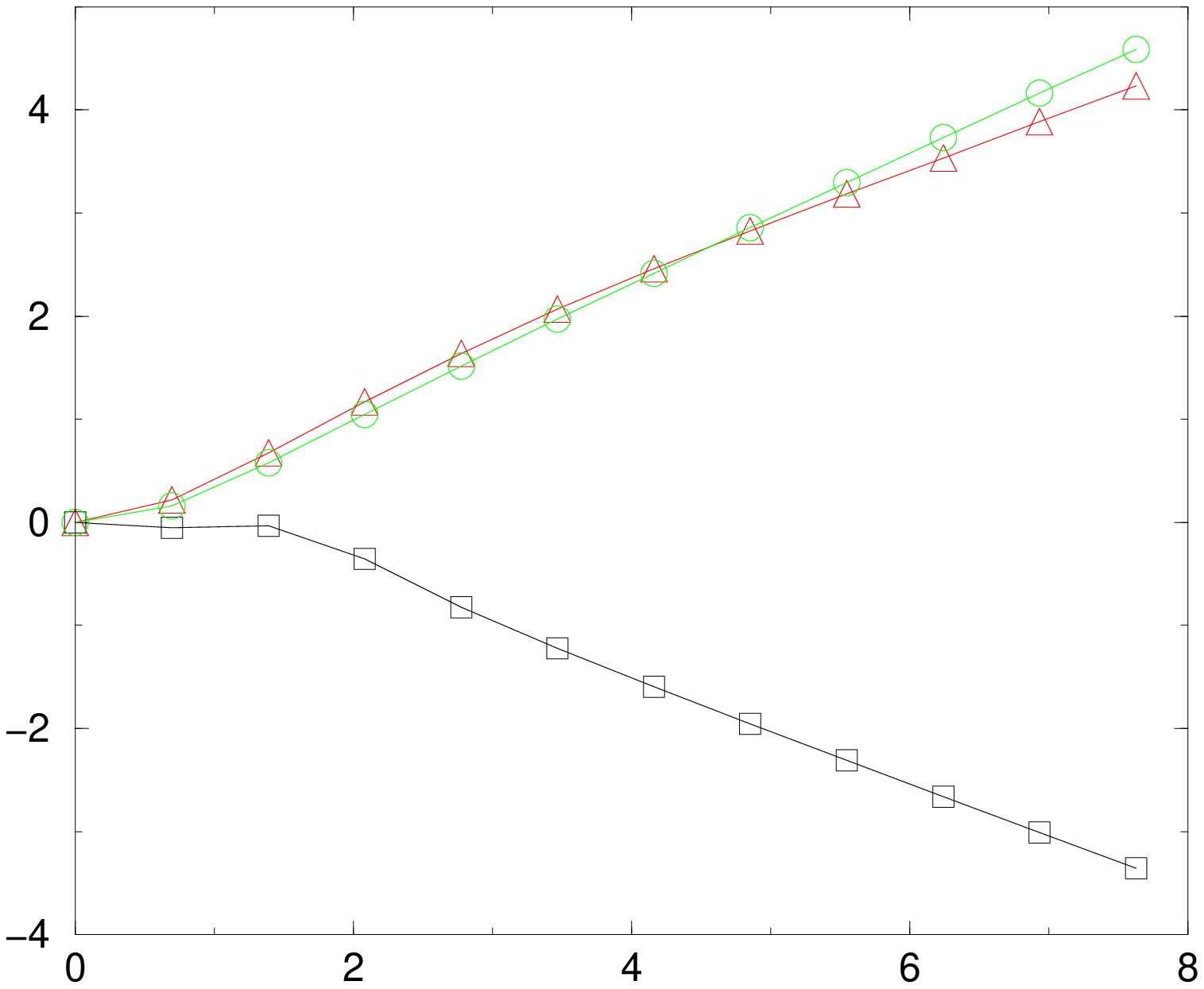}
\caption{ RG flow of the logarithm of the couplings in log-log scale \\
(a)  $Y_a = \ln (- \ln J_L^{typ}) \equiv \ln (- \overline{\ln J_L} )$ as a function of $X=\ln L$ : \\
(i) Ordered phase $\theta<\theta_c$ : the flow is non-monotonic, as shown here for $\theta=0.5$ (squares) and $\theta=1.245$ (diamond) : the asymptotic behavior (not visible on this plot appropriate to the critical region) corresponds to the classical random ferromagnet growth $J_L^{typ} \propto L^{d-1}=L $  \\
(ii) Disordered phase $\theta>\theta_c$ :   the asymptotic slope is $1$, as shown here for $\theta=1.5$ (triangles up) and $\theta=1.27$ (triangles right) \\
(iii) Critical point $\theta_c =1.256$ (circles) : the asymptotic slope is  $\psi \simeq 0.65 $. \\
(b)   $Y_b= \ln \Delta_{\ln J_L} \equiv \ln (\sqrt{ \overline{(\ln J_L)^2} -(\overline{\ln J_L})^2 })$ as a function of $X= \ln L$ : \\
(i) Ordered phase $\theta<\theta_c$ :   the asymptotic slope is $(-1/2)$, as shown here for $\theta=0.5$ (squares) \\
(ii) Disordered phase $\theta>\theta_c$ :   the asymptotic slope is $(+1/2)$, as shown here for $\theta=1.5$ (triangles up) \\
(iii) Critical point $\theta_c =1.256$ (circles) : the asymptotic slope is
  $\psi \simeq 0.65 $.
  }
\label{figflowj}
\end{figure}

On Fig. \ref{figflowj}, we show in log-log scale the RG flows of 
the typical renormalized coupling $J_L^{typ}$ of Eq. \ref{deftyp}
\begin{eqnarray}
\ln J_L^{typ} \vert_{\theta<\theta_c} && \oppropto_{L \to +\infty} \ln L  \nonumber \\
\ln J_L^{typ} \vert_{\theta=\theta_c}  && \oppropto_{L \to +\infty} - L^{\psi} \ \ {\rm with } \ \ \psi \simeq 0.65  \nonumber \\
\ln J_L^{typ} \vert_{\theta>\theta_c} && \oppropto_{L \to +\infty} -L 
\label{jtypflow}
\end{eqnarray}
and of the width $\Delta_{\ln J_L} $ of Eq. \ref{defwidth}
\begin{eqnarray}
\Delta_{\ln J_L} \vert_{\theta<\theta_c} && \oppropto_{L \to +\infty} L^{-\frac{1}{2}}   \nonumber \\
\Delta_{\ln J_L}  \vert_{\theta=\theta_c} && \oppropto_{L \to +\infty}  L^{\psi} \ \ {\rm with } \ \ \psi \simeq 0.65  \nonumber \\
\Delta_{\ln J_L}  \vert_{\theta>\theta_c}  && \oppropto_{L \to +\infty} L^{\frac{1}{2}} 
\label{jwidthflow}
\end{eqnarray}

At the critical point $\theta_c \simeq 1.256$, 
the RG flows of the typical value and of the width 
display the same activated scaling of Infinite Disorder Fixed Point
as in Eq. \ref{hLcriti}
\begin{eqnarray}
\ln J_L && = -  L^{\psi} u_c
\label{jLcriti}
\end{eqnarray}
where $u_c$ is some $O(1)$ random variable.

In the disordered phase, the typical renormalized coupling $J_{L}^{typ}$ decays
  exponentially with the size $L$
\begin{eqnarray}
\ln J_L^{typ} \equiv \overline{\ln J_L} \oppropto_{L \to +\infty} - \frac{L}{\xi_{typ}} 
\label{JLdisordertyp}
\end{eqnarray}
where $\xi_{typ} $ represents the typical correlation length.
 From our numerical data concerning the disordered phase, 
 the asymptotic behavior of Eq. \ref{JLdisordertyp} allows to measure $\xi_{typ}$ and its divergence
near criticality 
\begin{eqnarray}
\xi_{typ} \propto_{\theta \to \theta_c} (\theta_c-\theta)^{- \nu_{typ}} \ \ \ {\rm with } \ \ \ \nu_{typ}
 \simeq 0.44
\label{nutyp}
\end{eqnarray}

 The compatibility with the finite-size scaling form analogous to Eq. \ref{fsshtyp}
\begin{eqnarray}
\ln J_L^{typ} \equiv \overline{ \ln J_L }  = -  L^{\psi} F_J\left( L^{\frac{1}{\nu_{FS}}} ( \theta-\theta_c ) \right)
\label{fssjtyp}
\end{eqnarray}
 implies the relation
\begin{eqnarray}
\nu_{typ} = (1-\psi) \nu_{FS}
\label{nutypnufss}
\end{eqnarray}
It is satisfied by the previously quoted estimates of $\nu_{typ}$, $ \nu_{FS}$ and $\psi$.

\section{ Application to the two-dimensional spin-glass quantum Ising model }

\label{sec_d2spinglass}

To study the two-dimensional spin-glass quantum Ising model,
we have replaced the probability distribution of Eq. \ref{pJflat}
concerning the random ferromagnetic case by the flat distribution of zero-mean
\begin{eqnarray}
P_{SG}(J) = \frac{1}{2} \theta(-1 \leq J \leq 1)
\label{pJflatsg}
\end{eqnarray}
We have then repeated exactly the same numerical analysis as in 
the previous section \ref{sec_d2randomferro} : the critical point is now found at $\theta_c \simeq 1.088$, and the critical exponents $\psi \simeq 0.65$, $\nu_h \simeq 0.84$, $\nu_{typ} \simeq 0.44$, $\nu_{FS} \simeq 1.25$ are the same as for the random ferromagnet presented in the previous section, as expected for Infinite Disorder Fixed Points 
\cite{fisherreview,motrunich}, and as found within the Miyasaki-Nishimori asymmetric scheme \cite{nishiRandom}.

\section{ Conclusion }

\label{sec_conclusion}

In this paper, we have proposed a simple generalization in $d>1$ of the self-dual block renormalization procedure of Fernandez-Pacheco \cite{pacheco}, that we have tested for pure and random quantum Ising models, with the following conclusions.

For the pure models, where the Fernandez-Pacheco procedure is known to reproduce
the exact correlation length exponent $\nu(d=1)=1$ \cite{pacheco},
we have obtained $\nu (d=2) \simeq 0.625$  (to be compared with the 3D classical Ising model exponent $\nu \simeq 0.63$) and $\nu (d=3) \simeq 0.5018$ (to be compared with the 4D classical Ising model mean-field exponent $\nu =1/2$). 

For the random models, where the Fernandez-Pacheco procedure is known
to reproduce exactly the location of the critical point and
the critical exponents $\psi(=1)=1/2$, $\nu_{typ}=1$ and $\nu_{FS}=2$ of the Infinite Disorder Fixed Point \cite{nishiRandom}, we have applied numerically the renormalization rules to two-dimensional samples of linear size $L=4096 $, with 
either random ferromagnetic disorder or spin-glass disorder, both types of disorder leading to the same Infinite Disorder Fixed Point :
the finite-size correlation exponent $\nu_{FS} \simeq 1.25$ coincides with 
Strong Disorder Renormalization result (see \cite{kovacs2d,kovacsreview} and references therein), and with the asymmetric block renormalization of Ref \cite{nishiRandom},
but the activated exponent  $\psi \simeq 0.65$ turns out to be somewhat higher than
Strong Disorder Renormalization estimate $\psi \simeq 0.48$ (see \cite{kovacs2d,kovacsreview} and references therein). The origin of this difference remains to be clarified.
We have also analyzed the RG flows in the disordered and ordered phases, in order to extract the typical correlation length exponent $\nu_{typ} \simeq 0.44$ and 
the analog $\nu_h \simeq 0.84$, and tested the finite-size scaling.

In summary, the generalization in $d>1$ of the self-dual block renormalization procedure of Fernandez-Pacheco \cite{pacheco} is able to reproduce both the conventional scaling of pure critical points and the activated scaling of Infinite Disorder Fixed Points.
It would be thus interesting to develop such methods in models governed by Strong (not Infinite) Disorder Fixed Points like the Quantum Ising model with long-ranged interactions \cite{RGlong} or the superfluid-insulator transition \cite{refael}, as well as in models where the transition at weak disorder could be in another universality class
\cite{refael,majumdar}.

\appendix

\section{  Renormalization Rules in $d=3$  }

\label{sec_d3}

The initial quantum Hamiltonian on the cubic lattice reads
\begin{eqnarray}
H= - \sum_{(i,j,k)}  h (i,j,k) \sigma_{(i,j,k)}^x  
- \sum_{(i,j,k)} \sigma_{(i,j,k)}^z 
\left[  J_{\vec x}(i,j,k) \sigma_{(i+1,j,k)}^z 
+  J_{\vec y}(i,j,k) \sigma_{(i,j+1,k)}^z 
+  J_{\vec z}(i,j,k) \sigma_{(i,j,k+1)}^z  \right]
\label{h3d}
\end{eqnarray}

We wish to define a block renormalization rule, where each block of $2^3=8$ spins
($\sigma_{2i,2j,2k}$; $\sigma_{2i-1,2j,2k}$; $\sigma_{2i,2j-1,2k}$; $\sigma_{2i,2j,2k-1}$;
$\sigma_{2i-1,2j-1,2k}$; $\sigma_{2i-1,2j,2k-1}$; $\sigma_{2i,2j-1,2k-1}$;
$\sigma_{2i-1,2j-1,2k-1}$) will be replaced
by a single renormalized spin $(\sigma_{2i,2j,2k}^{RRR})$, after three elementary 
renormalization steps.

It is thus convenient to start by rewriting Eq. \ref{h3d} as
\begin{eqnarray}
&& H  =  \sum_{(i,j,k)} H_{i,j,k} 
\nonumber \\
&& H_{i,j,k} =  - h (2i,2j,2k) \sigma_{(2i,2j,2k)}^x 
- h (2i-1,2j-1,2k-1) \sigma_{(2i-1,2j-1,2k-1)}^x
\nonumber \\
&&  -  h (2i-1,2j,2k) \sigma_{(2i-1,2j,2k)}^x 
-  h (2i,2j-1,2k) \sigma_{(2i,2j-1,2k)}^x
-  h (2i,2j,2k-1) \sigma_{(2i,2j,2k-1)}^x
\nonumber \\
&&  -  h (2i-1,2j-1,2k) \sigma_{(2i-1,2j-1,2k)}^x
 -  h (2i,2j-1,2k-1) \sigma_{(2i,2j-1,2k-1)}^x
-  h (2i-1,2j,2k-1) \sigma_{(2i-1,2j,2k-1)}^x
\nonumber \\
&&  -   \sigma_{(2i-1,2j,2k)}^z 
[  J_{\vec x}(2i-1,2j,2k) \sigma_{(2i,2j,2k) }^z 
+  J_{\vec y}(2i-1,2j,2k) \sigma_{(2i-1,2j+1,2k)}^z 
+  J_{\vec z}(2i-1,2j,2k) \sigma_{(2i-1,2j,2k+1)}^z 
\nonumber \\
&& + J_{\vec x}(2i-2,2j,2k) \sigma_{(2i-2,2j,2k)}^z
 + J_{\vec y}(2i-1,2j-1,2k) \sigma_{(2i-1,2j-1,2k)}^z
 + J_{\vec z}(2i-1,2j,2k-1) \sigma_{(2i-1,2j,2k-1)}^z
 ]
\nonumber \\
&&  -   \sigma_{(2i,2j-1,2k)}^z 
[  J_{\vec y}(2i,2j-1,2k) \sigma_{(2i,2j,2k) }^z 
+  J_{\vec x}(2i,2j-1,2k) \sigma_{(2i+1,2j-1,2k)}^z
 +  J_{\vec z}(2i,2j-1,2k) \sigma_{(2i,2j-1,2k+1)}^z 
\nonumber \\
&& + J_{\vec y}(2i,2j-2,2k) \sigma_{(2i,2j-2,2k)}^z
 + J_{\vec x}(2i-1,2j-1,2k) \sigma_{(2i-1,2j-1,2k)}^z
 + J_{\vec z}(2i,2j-1,2k-1) \sigma_{(2i,2j-1,2k-1)}^z
 ]
\nonumber \\
&&  -   \sigma_{(2i,2j,2k-1)}^z 
[  J_{\vec z}(2i,2j,2k-1) \sigma_{(2i,2j,2k) }^z 
+  J_{\vec x}(2i,2j,2k-1) \sigma_{(2i+1,2j,2k-1)}^z
 +  J_{\vec y}(2i,2j,2k-1) \sigma_{(2i,2j+1,2k-1)}^z 
\nonumber \\
&& + J_{\vec z}(2i,2j,2k-2) \sigma_{(2i,2j,2k-2)}^z 
+ J_{\vec x}(2i-1,2j,2k-1) \sigma_{(2i-1,2j,2k-1)}^z
 + J_{\vec y}(2i,2j-1,2k-1) \sigma_{(2i,2j-1,2k-1)}^z
 ]
\nonumber \\
&&  -   \sigma_{(2i-1,2j-1,2k-1)}^z 
[  J_{\vec x}(2i-1,2j-1,2k-1) \sigma_{(2i-1,2j-1,2k) }^z 
+  J_{\vec y}(2i-1,2j-1,2k-1) \sigma_{(2i-1,2j,2k-1)}^z
\nonumber \\
&& +  J_{\vec z}(2i-1,2j-1,2k-1) \sigma_{(2i-1,2j-1,2k)}^z 
 + J_{\vec x}(2i-2,2j-1,2k-1) \sigma_{(2i-2,2j-1,2k-1)}^z 
\nonumber \\
&&+ J_{\vec y}(2i-1,2j-2,2k-1) \sigma_{(2i-1,2j-2,2k-1)}^z
 + J_{\vec z}(2i-1,2j-1,2k-2) \sigma_{(2i-1,2j-1,2k-2)}^z
 ]
\label{h3dbox}
\end{eqnarray}

\subsection{ First renormalization step }

In the first renormalization step, we choose the following intra-block Hamiltonian
\begin{eqnarray}
H_{intra}^{(1)} \equiv 
- h (2i-1,2j,2k) \sigma_{2i-1,2j,2k}^x 
-  J_{\vec x} (2i-1,2j,2k) \sigma_{2i-1,2j,2k}^z \sigma_{2i,2j,2k}^z 
\nonumber \\
- h (2i,2j-1,2k) \sigma_{2i,2j-1,2k}^x 
-  J_{\vec y}(2i,2j-1,2k)   \sigma_{2i,2j-1,2k}^z \sigma_{2i,2j,2k}^z 
\nonumber \\
- h (2i,2j,2k-1) \sigma_{2i,2j,2k-1}^x 
-  J_{\vec z}(2i,2j,2k-1)   \sigma_{2i,2j,2k-1}^z \sigma_{2i,2j,2k}^z 
\label{h1intra3d}
\end{eqnarray}
It has the form the Hamiltonian of Eq. \ref{hintra} analyzed in section \ref{sec_rgrule},
so that the four spins ( $\sigma_{2i,2j,2k}$;
$ \sigma_{2i-1,2j,2k}$ ; $\sigma_{2i,2j-1,2k}$ ;$ \sigma_{2i,2j,2k-1}$ )  are replaced by a single renormalized spin $(\sigma_{2i,2j,2k}^{R})$, whereas the four other spins $ \sigma_{2i-1,2j-1,2k};\sigma_{2i-1,2j,2k-1};\sigma_{2i,2j-1,2k-1};
\sigma_{2i-1,2j-1,2k-1}$
that are not involved in $H_{intra}^{(1)} $ remain unchanged.

The application of the projection rules of Eqs \ref{projsigmaz0}, \ref{projsigmazi}
and \ref{projsigmax0} read for the present case
\begin{eqnarray}
P_{intra}^{(1)} \sigma^z_{2i,2j,2k} P_{intra}^{(1)} && =  \sigma^z_{R(2i,2j,2k)}
\nonumber \\
P_{intra}^{(1)} \sigma^z_{2i-1,2j,2k}  P_{intra}^{(1)} && =
 \frac{ J_x(2i-1,2j,2k)}{ \sqrt{ h^2(2i-1,2j,2k)+J^2_{\vec x}(2i-1,2,2k) }}
  \sigma^z_{R(2i,2j,2k)}
\nonumber \\
P_{intra}^{(1)} \sigma^z_{2i,2j-1,2k}  P_{intra}^{(1)} && =
 \frac{ J_y(2i,2j-1,2k)}{ \sqrt{ h^2(2i,2j-1,2k)+J^2_{\vec y}(2i,2j-1,2k) }} 
 \sigma^z_{R(2i,2j,2k)}
\nonumber \\
P_{intra}^{(1)} \sigma^z_{2i,2j,2k-1}  P_{intra}^{(1)} && =
 \frac{ J_z(2i,2j,2k-1)}{ \sqrt{ h^2(2i,2j,2k-1)+J^2_{\vec z}(2i,2j,2k-1) }} 
 \sigma^z_{R(2i,2j,2k)}
\nonumber \\
P_{intra}^{(1)} \sigma^x_{2i,2j,2k} P_{intra}^{(1)} && =
 \frac{h (2i-1,2j,2k)}{\sqrt{h^2(2i-1,2j,2k)+J^2_{\vec x}(2i-1,2j,2k) } }
\frac{h (2i,2j-1,2k)}{\sqrt{h^2(2i,2j-1,2k)+J^2_{\vec y}(2i,2j-1,2k) } }
\nonumber \\
&& \frac{h (2i,2j,2k-1)}{\sqrt{h^2(2i,2j,2k-1)+J^2_{\vec y}(2i,2j,2k-1) } }
 \sigma_{R(2i,2j,2k)}^x
\label{proj3dp1}
\end{eqnarray}

As a consequence, the projection of the remaining part of the Hamiltonian reads
\begin{eqnarray}
&& H^R  = P_{intra}^{(1)} \left[ \sum_{(i,j,k)} (H_{i,j,k}-H_{i,j,k}^{(1)} )  \right] P_{intra}^{(1)} =  
\sum_{(i,j)} H_{i,j,k}^{R} 
\nonumber \\
&& H_{i,j,k}^{R} = 
  - h^R (2i,2j,2k) \sigma_{R(2i,2j,2k)}^x - h (2i-1,2j-1,2k-1) \sigma_{(2i-1,2j-1,2k-1)}^x
\nonumber \\
&&  -  h (2i-1,2j-1,2k) \sigma_{(2i-1,2j-1,2k)}^x -  h (2i,2j-1,2k-1) \sigma_{(2i,2j-1,2k-1)}^x
-  h (2i-1,2j,2k-1) \sigma_{(2i-1,2j,2k-1)}^x
\nonumber \\
&&  - \sigma^z_{R(2i,2j,2k)}
[ J_{2\vec x}^R(2i-2,2j,2k) \sigma_{R(2i-2,2j,2k)}^z  
 + J_{2 \vec y}^R(2i,2j-2,2k) \sigma_{R(2i,2j-2,2k)}^z
+ J_{2\vec z}^R(2i,2j,2k-2) \sigma_{R(2i,2j,2k-2)}^z 
\nonumber \\
&&
+ J^R_{ \vec x- \vec y} (2i,2j,2k)\sigma_{2i+1,2j-1,2k}^z
+ J^R_{ -\vec x+ \vec y} (2i,2j,2k) \sigma_{2i-1,2j+1,2k}^z
\nonumber \\
&& + J^R_{ \vec x- \vec z} (2i,2j,2k) \sigma_{2i+1,2j,2k-1}^z
+ J^R_{ -\vec x+ \vec z} (2i,2j,2k)  \sigma_{2i-1,2j,2k+1}^z
\nonumber \\
&&
+J^R_{ \vec y- \vec z} (2i,2j,2k)\sigma_{2i,2j+1,2k-1}^z 
+J^R_{ -\vec y+ \vec z} (2i,2j,2k)\sigma_{2i,2j-1,2k+1}^z 
\nonumber \\
&& 
+ J^R_{ \vec x+ \vec y} (2i-1,2j-1,2k) \sigma_{2i-1,2j-1,2k}^z
+ J^R_{ \vec x+ \vec z} (2i-1,2j,2k-1) \sigma_{2i-1,2j,2k-1}^z
\nonumber \\
&& + J^R_{ \vec y+ \vec z} (2i,2j-1,2k-1) \sigma_{2i,2j-1,2k-1}^z
]
\nonumber \\
&&  -   \sigma_{(2i-1,2j-1,2k-1)}^z 
[  J_{\vec x}(2i-1,2j-1,2k-1) \sigma_{(2i-1,2j-1,2k) }^z 
+  J_{\vec y}(2i-1,2j-1,2k-1) \sigma_{2i-1,2j,2k-1}^z
\nonumber \\
&& +  J_{\vec z}(2i-1,2j-1,2k-1) \sigma_{2i-1,2j-1,2k}^z 
 + J_{\vec x}(2i-2,2j-1,2k-1) \sigma_{(2i-2,2j-1,2k-1)}^z 
\nonumber \\
&&+ J_{\vec y}(2i-1,2j-2,2k-1) \sigma_{(2i-1,2j-2,2k-1)}^z
 + J_{\vec z}(2i-1,2j-1,2k-2) \sigma_{(2i-1,2j-1,2k-2)}^z
 ]
\label{h3dRfin}
\end{eqnarray}
with

(i)  the renormalized transverse fields
\begin{eqnarray}
h^R (2i,2j,2k) && =h(2i,2j,2k) 
\frac{h (2i-1,2j,2k)}{\sqrt{h^2(2i-1,2j,2k)+J^2_{\vec x}(2i-1,2j,2k) } }
\nonumber \\ && \frac{h (2i,2j-1,2k)}{\sqrt{h^2(2i,2j-1,2k)+J^2_{\vec y}(2i,2j-1,2k) } }
\frac{h (2i,2j,2k-1)}{\sqrt{h^2(2i,2j,2k-1)+J^2_{\vec z}(2i,2j,2k-1) } }
\label{rght3d}
\end{eqnarray}

(ii) the renormalized couplings along the lattice directions at distance two
\begin{eqnarray}
J^R_{2 \vec x} (2i-2,2j,2k) && = J_{ \vec x}(2i-2,2j,2k) 
\frac{J_{ \vec x} (2i-1,2j,2k)}{\sqrt{h^2(2i-1,2j,2k)+J^2_{ \vec x}(2i-1,2j,2k) } }
\nonumber \\
J^R_{2 \vec y} (2i,2j-2,2k) && = J_{ \vec y}(2i,2j-2,2k) 
\frac{J_{ \vec y} (2i,2j-1,2k)}{\sqrt{h^2(2i,2j-1,2k)+J^2_{ \vec y}(2i,2j-1,2k) } }
\nonumber \\
J^R_{2 \vec z} (2i,2j,2k-2) && = J_{ \vec z}(2i,2j,2k-2) 
\frac{J_{ \vec z} (2i,2j,2k-1)}{\sqrt{h^2(2i,2j,2k-1)+J^2_{ \vec z}(2i,2j,2k-1) } }
\label{rgj3d}
\end{eqnarray}

(iii) the renormalized couplings along the diagonal directions $(\vec x- \vec y)$ ; $(-\vec x+ \vec y)$; $(\vec x- \vec z)$ ; $(-\vec x+ \vec z)$ ; $(\vec y- \vec z)$ ; $(-\vec y+ \vec z)$
\begin{eqnarray}
J^R_{ \vec x- \vec y} (2i,2j,2k) && = J_{ \vec x}(2i,2j-1,2k) 
\frac{J_{ \vec y} (2i,2j-1,2k)}{\sqrt{h^2(2i,2j-1,2k)+J^2_{ \vec y}(2i,2j-1,2k) } }
\nonumber \\
J^R_{ -\vec x+ \vec y} (2i,2j,2k) && = J_{ \vec y}(2i-1,2j,2k) 
\frac{J_{ \vec x} (2i-1,2j,2k)}{\sqrt{h^2(2i-1,2j,2k)+J^2_{ \vec x}(2i-1,2j,2k) } }
\nonumber \\
J^R_{ \vec x- \vec z} (2i,2j,2k) && = J_{ \vec x}(2i,2j,2k-1) 
\frac{J_{ \vec z} (2i,2j,2k-1)}{\sqrt{h^2(2i,2j,2k-1)+J^2_{ \vec y}(2i,2j,2k-1) } }
\nonumber \\
J^R_{ -\vec x+ \vec z} (2i,2j,2k) && = J_{ \vec z}(2i-1,2j,2k) 
\frac{J_{ \vec x} (2i-1,2j,2k)}{\sqrt{h^2(2i-1,2j,2k)+J^2_{ \vec x}(2i-1,2j,2k) } }
\nonumber \\
J^R_{ \vec y- \vec z} (2i,2j,2k) && = J_{ \vec y}(2i,2j,2k-1) 
\frac{J_{ \vec z} (2i,2j,2k-1)}{\sqrt{h^2(2i,2j,2k-1)+J^2_{ \vec z}(2i,2j,2k-1) } }
\nonumber \\
J^R_{ -\vec y+ \vec z} (2i,2j,2k) && = J_{ \vec z}(2i,2j-1,2k) 
\frac{J_{ \vec y} (2i,2j-1,2k)}{\sqrt{h^2(2i,2j-1,2k)+J^2_{ \vec y}(2i,2j-1,2k) } }
\label{rgj3ddiagexter}
\end{eqnarray}

(iv) the renormalized couplings along the diagonal directions $(\vec x+ \vec y)$; $(\vec x+ \vec z)$ ; $(\vec y+ \vec z)$
\begin{eqnarray}
J^R_{ \vec x+ \vec y} (2i-1,2j-1,2k) && = J_{ \vec x}(2i-1,2j-1,2k) 
\frac{J_{ \vec y} (2i,2j-1,2k)}{\sqrt{h^2(2i,2j-1,2k)+J^2_{ \vec y}(2i,2j-1,2k) } }
\nonumber \\
&&+  J_{ \vec y}(2i-1,2j-1,2k) 
\frac{J_{ \vec x} (2i-1,2j,2k)}{\sqrt{h^2(2i-1,2j,2k)+J^2_{ \vec x}(2i-1,2j,2k) } }
\nonumber \\
J^R_{ \vec x+ \vec z} (2i-1,2j,2k-1) && = J_{ \vec x}(2i-1,2j,2k-1) 
\frac{J_{ \vec z} (2i,2j,2k-1)}{\sqrt{h^2(2i,2j,2k-1)+J^2_{ \vec z}(2i,2j,2k-1) } }
\nonumber \\
&&+  J_{ \vec z}(2i-1,2j,2k-1) 
\frac{J_{ \vec x} (2i-1,2j,2k)}{\sqrt{h^2(2i-1,2j,2k)+J^2_{ \vec x}(2i-1,2j,2k) } }
\nonumber \\
J^R_{ \vec y+ \vec z} (2i,2j-1,2k-1) && = J_{ \vec y}(2i,2j-1,2k-1) 
\frac{J_{ \vec z} (2i,2j,2k-1)}{\sqrt{h^2(2i,2j,2k-1)+J^2_{ \vec z}(2i,2j,2k-1) } }
\nonumber \\
&&+  J_{ \vec z}(2i,2j-1,2k-1) 
\frac{J_{ \vec y} (2i,2j-1,2k)}{\sqrt{h^2(2i,2j-1,2k)+J^2_{ \vec y}(2i,2j-1,2k) } }
\label{rgj3ddiagyzinter}
\end{eqnarray}

\subsection{ Second renormalization step }

For the second renormalization step, we choose the following intra-block Hamiltonian
\begin{eqnarray}
H_{intra}^{(2)} \equiv 
&& - h (2i-1,2j-1,2k) \sigma_{(2i-1,2j-1,2k)}^x -  J^R_{\vec x+\vec y} (2i-1,2j-1,2k)
  \sigma_{(2i-1,2j-1,2k)}^z \sigma^z_{R(2i,2j,2k)}
\nonumber \\
&& - h (2i-1,2j,2k-1) \sigma_{(2i-1,2j,2k-1)}^x -  J^R_{\vec x+\vec z} (2i-1,2j,2k-1)
 \sigma_{(2i-1,2j,2k-1)}^z \sigma^z_{R(2i,2j,2k)}
\nonumber \\
&& - h (2i,2j-1,2k-1) \sigma_{(2i,2j-1,2k-1)}^x -  J^R_{\vec y+\vec z} (2i,2j-1,2k-1)
  \sigma_{(2i,2j-1,2k-1)}^z \sigma^z_{R(2i,2j,2k)}
\label{h2intra3d}
\end{eqnarray}
It has the form the Hamiltonian of Eq. \ref{hintra} analyzed in section \ref{sec_rgrule},
so that the four spins ($\sigma^R_{(2i,2j,2k)}$ ;$ \sigma_{(2i-1,2j-1,2k)}$ ;$ \sigma_{(2i-1,2j,2k-1)}$ ;
$ \sigma_{(2i,2j-1,2k-1)}$) are replaced by a single renormalized spin $\sigma_{RR(2i,2j,2k)}$.

The application of the projection rules of Eqs \ref{projsigmaz0}, \ref{projsigmazi}
and \ref{projsigmax0} read for the present case
\begin{eqnarray}
&& P_{intra}^{(2)} \sigma^z_{R(2i,2j,2k)} P_{intra}^{(2)} =  \sigma^z_{RR(2i,2j,2k)}
\label{proj3dp2}
\\
&& P_{intra}^{(2)} \sigma^z_{2i-1,2j-1,2k}  P_{intra}^{(2)} =
 \frac{ J^R_{\vec x+\vec y} (2i-1,2j-1,2k) }
{ \sqrt{ h^2(2i-1,2j-1,2k)+[J^R_{\vec x+\vec y} (2i-1,2j-1,2k)]^2 }}  \sigma^z_{RR(2i,2j,2k)}
\nonumber \\
&& P_{intra}^{(2)} \sigma^z_{2i-1,2j,2k-1}  P_{intra}^{(2)} =
 \frac{ J^R_{\vec x+\vec z} (2i-1,2j,2k-1) }
{ \sqrt{ h^2(2i-1,2j,2k-1)+[J^R_{\vec x+\vec z} (2i-1,2j,2k-1)]^2 }}  \sigma^z_{RR(2i,2j,2k)}
\nonumber \\
&& P_{intra}^{(2)} \sigma^z_{2i,2j-1,2k-1}  P_{intra}^{(2)} =
 \frac{ J^R_{\vec y+\vec z} (2i,2j-1,2k-1) }
{ \sqrt{ h^2(2i,2j-1,2k-1)+[J^R_{\vec y+\vec z} (2i,2j-1,2k-1)]^2 }}  \sigma^z_{RR(2i,2j,2k)}
\nonumber \\
&& P_{intra}^{(2)} \sigma^x_{R(2i,2j,2k)} P_{intra}^{(2)} =
 \frac{h (2i-1,2j-1,2k)}{\sqrt{h^2(2i-1,2j-1,2k)+[J^R_{\vec x+\vec y} (2i-1,2j-1,2k)]^2 } }
\nonumber \\
&& \frac{h (2i-1,2j,2k-1)}{\sqrt{h^2(2i-1,2j,2k-1)+[J^R_{\vec x+\vec z} (2i-1,2j,2k-1)]^2 } }
  \frac{h (2i,2j-1,2k-1)}{\sqrt{h^2(2i,2j-1,2k-1)+[J^R_{\vec y+\vec z} (2i,2j-1,2k-1)]^2 } }
 \sigma_{RR(2i,2j)}^x
\nonumber
\end{eqnarray}

As a consequence, the projection of the remaining part of the Hamiltonian reads
\begin{eqnarray}
&& H^{RR}  = P_{intra}^{(2)} \left[ \sum_{(i,j,k)} (H_{i,j,k}^R-H_{i,j,k}^{(2)} )  \right] P_{intra}^{(2)} =  
\sum_{(i,j)} H_{i,j,k}^{RR} 
\nonumber \\
&& H_{i,j,k}^{RR} =   
 - h^{RR} (2i,2j,2k) \sigma_{RR(2i,2j,2k)}^x - h (2i-1,2j-1,2k-1) \sigma_{(2i-1,2j-1,2k-1)}^x
\nonumber \\
&& - J_{2\vec x}^{RR}(2i-2,2j,2k) \sigma_{RR(2i-2,2j,2k)}^z  \sigma^z_{RR(2i,2j,2k)}
 - J_{2 \vec y}^{RR}(2i,2j-2,2k) \sigma_{RR(2i,2j-2,2k)}^z \sigma^z_{RR(2i,2j,2k)}
\nonumber \\
&& - J_{2\vec z}^{RR}(2i,2j,2k-2) \sigma_{RR(2i,2j,2k-2)}^z  \sigma^z_{RR(2i,2j,2k)}
\nonumber \\
&& - J^{RR}_{\vec x-\vec y-\vec z} (2i,2j,2k) \sigma^z_{RR(2i,2j,2k)} \sigma_{(2i+1,2j-1,2k-1)}^z 
-J^{RR}_{-\vec x+\vec y-\vec z} (2i,2j,2k)  \sigma^z_{RR(2i,2j,2k)}  \sigma_{(2i-1,2j+1,2k-1)}^z 
\nonumber \\
&& - J^{RR}_{-\vec x-\vec y+\vec z} (2i,2j,2k) \sigma^z_{RR(2i,2j,2k)} \sigma_{(2i-1,2j-1,2k+1)}^z 
\nonumber \\
&&  -  J^{RR}_{\vec x+\vec y+\vec z} (2i-1,2j-1,2k-1) \sigma_{(2i-1,2j-1,2k-1)}^z \sigma^z_{RR(2i,2j,2k)}
\label{h3dRRfin}
\end{eqnarray}
in terms of

(i)  the renormalized transverse fields
\begin{eqnarray}
&& h^{RR} (2i,2j,2k)  =h^R(2i,2j,2k) 
\frac{h (2i-1,2j-1,2k)}{\sqrt{h^2(2i-1,2j-1,2k)+[J^R_{\vec x+\vec y}(2i-1,2j-1,2k)]^2 } }
\nonumber \\
&& 
\frac{h (2i-1,2j,2k-1)}{\sqrt{h^2(2i-1,2j,2k-1)+[J^R_{\vec x+\vec z}(2i-1,2j,2k-1)]^2 } }
\frac{h (2i,2j-1,2k-1)}{\sqrt{h^2(2i,2j-1,2k-1)+[J^R_{\vec y+\vec z}(2i,2j-1,2k-1)]^2 } }
\label{rghRR3d}
\end{eqnarray}

(ii) the renormalized couplings along the lattice directions at distance two
\begin{eqnarray}
J^{RR}_{2 \vec x} (2i-2,2j,2k) && = J^R_{2 \vec x}(2i-2,2j,2k)
\nonumber \\ &&  + J^R_{ \vec x- \vec y} (2i-2,2j,2k)
\frac{J^R_{ \vec x+ \vec y} (2i-1,2j-1,2k)}{\sqrt{[h^R(2i-1,2j-1,2k)]^2
+[J^R_{ \vec x+ \vec y}(2i-1,2j-1,2k)]^2 } }
\nonumber \\ && 
+ J^R_{ \vec x- \vec z} (2i-2,2j,2k)
\frac{J^R_{ \vec x+ \vec z} (2i-1,2j,2k-1)}{\sqrt{[h^R(2i-1,2j,2k-1)]^2
+[J^R_{ \vec x+ \vec z}(2i-1,2j,2k-1)]^2 } }
\nonumber \\ 
J^{RR}_{2 \vec y} (2i,2j-2,2k) && = J^R_{2 \vec y}(2i,2j-2,2k)
\nonumber \\ &&  +J^R_{ -\vec x+ \vec y} (2i,2j-2,2k)
\frac{J^R_{ \vec x+ \vec y} (2i-1,2j-1,2k)}
{\sqrt{[h^R(2i-1,2j-1,2k)]^2+[J^R_{ \vec x+ \vec y}(2i-1,2j-1,2k)]^2 } }
\nonumber \\ && 
 + J^R_{ \vec y- \vec z} (2i,2j-2,2k)
\frac{J^R_{ \vec y+ \vec z} (2i,2j-1,2k-1)}
{\sqrt{[h^R(2i,2j-1,2k-1)]^2+[J^R_{ \vec y+ \vec z}(2i,2j-1,2k-1)]^2 } }
 \nonumber \\  
J^{RR}_{2 \vec z} (2i,2j,2k-2) && = J^R_{2 \vec z}(2i,2j,2k-2)
\nonumber \\ &&  + J^R_{ -\vec x+ \vec z} (2i,2j,2k-2) 
\frac{J^R_{ \vec x+ \vec z} (2i-1,2j,2k-1)}
{\sqrt{[h^R(2i-1,2j,2k-1)]^2+[J^R_{ \vec x+ \vec z}(2i-1,2j,2k-1)]^2 } }
\nonumber \\ && 
+ J^R_{ -\vec y+ \vec z} (2i,2j,2k-2)
\frac{J^R_{ \vec y+ \vec z} (2i,2j-1,2k-1)}
{\sqrt{[h^R(2i,2j-1,2k-1)]^2+[J^R_{ \vec y+ \vec z}(2i,2j-1,2k-1)]^2 } }
\label{rgj3ddRRdiagneg}
\end{eqnarray}

(iii) the renormalized couplings along $(\vec x-\vec y-\vec z)$ ; 
 $(-\vec x+\vec y-\vec z)$ ; $(-\vec x+\vec y-\vec z)$
\begin{eqnarray}
J^{RR}_{\vec x-\vec y-\vec z} (2i,2j,2k)
&& = J_{\vec x}(2i,2j-1,2k-1)  \frac{ J^R_{\vec y+\vec z} (2i,2j-1,2k-1) }
{ \sqrt{ h^2(2i,2j-1,2k-1)+[J^R_{\vec y+\vec z} (2i,2j-1,2k-1)]^2 }}
\nonumber \\
J^{RR}_{-\vec x+\vec y-\vec z} (2i,2j,2k)
&& = J_{\vec y}(2i-1,2j,2k-1) \frac{ J^R_{\vec x+\vec z} (2i-1,2j,2k-1) }
{ \sqrt{ h^2(2i-1,2j,2k-1)+[J^R_{\vec x+\vec z} (2i-1,2j,2k-1)]^2 }}
\nonumber \\
J^{RR}_{-\vec x-\vec y+\vec z} (2i,2j,2k)
&& = J_{\vec z}(2i-1,2j-1,2k)  \frac{ J^R_{\vec x+\vec y} (2i-1,2j-1,2k) }
{ \sqrt{ h^2(2i-1,2j-1,2k)+[J^R_{\vec x+\vec y} (2i-1,2j-1,2k)]^2 }} 
\label{proj3djrrdiagneg}
\end{eqnarray}

(iv) the renormalized couplings along $(\vec x+\vec y+\vec z)$
\begin{eqnarray}
J^{RR}_{\vec x+\vec y+\vec z} (2i-1,2j-1,2k-1)
&& =  J_{\vec x}(2i-1,2j-1,2k-1)  \frac{ J^R_{\vec y+\vec z} (2i,2j-1,2k-1) }
{ \sqrt{ h^2(2i,2j-1,2k-1)+[J^R_{\vec y+\vec z} (2i,2j-1,2k-1)]^2 }}
\nonumber \\
&& + J_{\vec y}(2i-1,2j-1,2k-1) \frac{ J^R_{\vec x+\vec z} (2i-1,2j,2k-1) }
{ \sqrt{ h^2(2i-1,2j,2k-1)+[J^R_{\vec x+\vec z} (2i-1,2j,2k-1)]^2 }}  
\nonumber \\
&& + J_{\vec z}(2i-1,2j-1,2k-1)\frac{ J^R_{\vec x+\vec y} (2i-1,2j-1,2k) }
{ \sqrt{ h^2(2i-1,2j-1,2k)+[J^R_{\vec x+\vec y} (2i-1,2j-1,2k)]^2 }}
\label{proj3drrdiagpos}
\end{eqnarray}

\subsection{ Third renormalization step }

For the third renormalization step, we choose the following intra-block Hamiltonian
\begin{eqnarray}
H_{i,j,k}^{(3)} \equiv 
- h (2i-1,2j-1,2k-1) \sigma_{(2i-1,2j-1,2k-1)}^x
 -  J^{RR}_{\vec x+\vec y+\vec z} (2i-1,2j-1,2k-1) \sigma_{(2i-1,2j-1,2k-1)}^z \sigma^z_{RR(2i,2j,2k)}
\label{h3intra3d}
\end{eqnarray}
It has the form the Hamiltonian of Eq. \ref{hintra} analyzed in section \ref{sec_rgrule},
so that the two spins $(\sigma_{(2i-1,2j-1,2k-1)}^z , \sigma^z_{RR(2i,2j,2k)})$ can be replaced by a single renormalized spin $(\sigma_{RRR(2i,2j,2k)})$.
The application of the projection rules of Eqs \ref{projsigmaz0}, \ref{projsigmazi}
and \ref{projsigmax0} read for the present case
\begin{eqnarray}
P_{intra}^{(3)} \sigma^z_{RR(2i,2j,2k)} P_{intra}^{(2)} && =  \sigma^z_{RRR(2i,2j,2k)}
\nonumber \\
P_{intra}^{(3)} \sigma^z_{2i-1,2j-1,2k-1}  P_{intra}^{(2)} && =
 \frac{ J^{RR}_{\vec x+\vec y+\vec z} (2i-1,2j-1,2k-1) }
{ \sqrt{ h^2(2i-1,2j-1,2k-1)+[J^{RR}_{\vec x+\vec y+\vec z} (2i-1,2j-1,2k-1)]^2 }}
  \sigma^z_{RRR(2i,2j,2k)}
\nonumber \\
P_{intra}^{(3)} \sigma^x_{RR(2i,2j,2k)} P_{intra}^{(2)} && =
 \frac{ h (2i-1,2j-1,2k-1) }
{ \sqrt{ h^2(2i-1,2j-1,2k-1)+[J^{RR}_{\vec x+\vec y+\vec z} (2i-1,2j-1,2k-1)]^2 }}
 \sigma_{RRR(2i,2j,2k)}^x
\label{proj3dp3}
\end{eqnarray}

The projection of the remaining part of the Hamiltonian reads
\begin{eqnarray}
H^{RRR} && = P_{intra}^{(3)} \left[ \sum_{(i,j,k)} (H_{i,j,k}^{RR}-H_{i,j,k}^{(3)} )  \right] P_{intra}^{(3)} =  
\sum_{(i,j,k)} H_{i,j,k}^{RRR} 
\nonumber \\
H_{i,j,k}^{RRR} = &&  
 - h^{RRR} (2i,2j,2k) \sigma_{RRR(2i,2j,2k)}^x 
\nonumber \\
&& - J_{2\vec x}^{RRR}(2i-2,2j,2k) \sigma_{RRR(2i-2,2j,2k)}^z  \sigma^z_{RRR(2i,2j,2k)}
 - J_{2 \vec y}^{RRR}(2i,2j-2,2k) \sigma_{RRR(2i,2j-2,2k)}^z \sigma^z_{RRR(2i,2j,2k)}
\nonumber \\
&& - J_{2\vec z}^{RRR}(2i,2j,2k-2) \sigma_{RRR(2i,2j,2k-2)}^z  \sigma^z_{RRR(2i,2j,2k)}
\label{h3dRRRfin}
\end{eqnarray}
i.e. it has the same form as the initial Hamiltonian, in terms of

(i)  the renormalized transverse fields 
\begin{eqnarray}
h^{RRR} (2i,2j,2k) && =h^{RR}(2i,2j,2k) 
\frac{ h (2i-1,2j-1,2k-1) }
{ \sqrt{ h^2(2i-1,2j-1,2k-1)+[J^{RR}_{\vec x+\vec y+\vec z} (2i-1,2j-1,2k-1)]^2 }}
\label{rghRRR}
\end{eqnarray}

(ii) the renormalized couplings along the lattice directions at distance two

\begin{eqnarray}
J^{RRR}_{2 \vec x} (2i-2,2j,2k) && = J^{RR}_{2 \vec x}(2i-2,2j,2k)
\nonumber \\
&& + J^{RR}_{\vec x-\vec y-\vec z} (2i-2,2j,2k)
 \frac{ J^{RR}_{\vec x+\vec y+\vec z} (2i-1,2j-1,2k-1) }
{ \sqrt{ h^2(2i-1,2j-1,2k-1)+[J^{RR}_{\vec x+\vec y+\vec z} (2i-1,2j-1,2k-1)]^2 }}
\nonumber \\ 
J^{RRR}_{2 \vec y} (2i,2j-2,2k) && = J^{RR}_{2 \vec y}(2i,2j-2,2k)
\nonumber \\
&&  +J^{RR}_{-\vec x+\vec y-\vec z} (2i,2j-2,2k)
 \frac{ J^{RR}_{\vec x+\vec y+\vec z} (2i-1,2j-1,2k-1) }
{ \sqrt{ h^2(2i-1,2j-1,2k-1)+[J^{RR}_{\vec x+\vec y+\vec z} (2i-1,2j-1,2k-1)]^2 }}
 \nonumber \\  
J^{RRR}_{2 \vec z} (2i,2j,2k-2) && = J^{RR}_{2 \vec z}(2i,2j,2k-2)
\nonumber \\
&&  + J^{RR}_{-\vec x-\vec y+\vec z} (2i,2j,2k-2)
 \frac{ J^{RR}_{\vec x+\vec y+\vec z} (2i-1,2j-1,2k-1) }
{ \sqrt{ h^2(2i-1,2j-1,2k-1)+[J^{RR}_{\vec x+\vec y+\vec z} (2i-1,2j-1,2k-1)]^2 }}
\label{rgj3ddRRRv}
\end{eqnarray}

\subsection{ Application to the pure quantum Ising model in $d=3$ }

If we start from the pure model of parameters $(h,J)$,
the renormalization rules of Eqs \ref{rghRRR} and \ref{rgj3ddRRRv} reads
for the ratio $K \equiv \frac{J}{h}$ to
\begin{eqnarray}
K^{RRR} && \equiv \frac{J^{RRR} }{h_{RRR} }
\nonumber \\
&& = K^2 \sqrt{1+K^2+4 K^4}
\left[ 12 K^4 + (4 K^2+\sqrt{1+K^2+4 K^4} ) \sqrt{ \frac{1+K^2+4 K^4+36 K^6}{1+K^2}}
\right] \equiv \phi(K)
\label{evolk3d}
\end{eqnarray}
The critical point satisfying $K_c=\phi(K_c)$ is found to be
\begin{eqnarray}
K_c \simeq 0.398425
\label{kc3d}
\end{eqnarray}
The correlation length exponent $\nu$ given by $2^{\frac{1}{\nu}}=\phi'(K_c)$
\begin{eqnarray}
\nu \simeq  0.5018
\label{nu3d}
\end{eqnarray}
is very close to the mean-field value $\nu_{MF}=1/2$ of the 4D classical Ising model.

\end{document}